\documentclass[]{aa} 

\usepackage{graphicx,amssymb,psfig,natbib}

\newcommand{\AaA}{A\&A}
\newcommand{\AaAS}{A\&As}

\newcommand{\ApJS}{Astrophysical Journal Supplement Series}
\newcommand{\AJ}{AJ}
\newcommand{\ApJ}{ApJ}
\newcommand{\MNRAS}{MNRAS}

\newcommand{\Sci}{Science}

\newcommand{\Msun} {M$_\odot$}
\newcommand{\Lsun} {L$_\odot$}
\newcommand{\Teff} {T$_{\rm{eff}}$}
\newcommand{\Tstar} {T$_{\rm{eff}}$}
\newcommand{\Lstar} {L$_\star$}
\newcommand{\Mstar} {M$_\star$}
\newcommand{\Rstar} {R$_\star$}

\newcommand{\um} {$\mu$m}

\newcommand{\Md} {M$_{d}$}
\newcommand{\Ri} {R$_{in}$}
\newcommand{\Rd} {R$_{d}$}
\newcommand{\Hp} {H$_p$}
\newcommand{\Hs} {H$_s$}

\newcommand{\cmc} {cm$^{-3}$}
\newcommand{\Wm} {W m$^{-2}$}

\newcommand{\simless}{\mathbin{\lower 3pt\hbox
      {$\rlap{\raise 5pt\hbox{$\char'074$}}\mathchar"7218$}}} 
\newcommand{\simgreat}{\mathbin{\lower 3pt\hbox
     {$\rlap{\raise 5pt\hbox{$\char'076$}}\mathchar"7218$}}} 

\begin{document}

\title{PAHs in Circumstellar Disks around Herbig Ae/Be stars}
\author{Emilie Habart\inst{1},
Antonella Natta\inst{1}, 
Endrik Kr\"ugel\inst{2}}

\institute{ 
    Osservatorio Astrofisico di Arcetri, INAF, Largo E.Fermi 5,
    I-50125 Firenze, Italy 
\and
    Max-Planck Instute fur Radioastronomie, Auf dem H\"ugel 69, Bonn, Germany}

\abstract{We investigate the presence and properties of PAHs on the surface of
circumstellar disks around Herbig Ae/Be stars by comparing the predictions of
disk models with observations.  We present results of a radiation transfer code
for disks heated by the central star, in hydrostatic equilibrium in the vertical
direction (flared disks).  The dust is a mixture of large grains in thermal
equilibrium, transiently heated small grains and PAHs.  Special attention is
given to the influence of the stellar, disk and PAHs properties on the strength
of the PAH emission lines and their spatial distribution.  The models predict an
infrared SED showing PAH features at 3.3, 6.2, 7.7, and 11.3 \um\ clearly visible
above the continuum, and with some of them very strong.  The PAH emission,
spatially extended, comes mostly from the outer disk region ($R\sim$100 AU)
while the continuum emission at similar wavelengths, mostly due to warm large
grains, is confined to the innermost disk regions ($R\sim$few AU).  We compare
the model results to infrared observations from ISO and ground-based telescopes
of some thirty Herbig Ae/Be stars.  Most of the observed PAH features in objects
with spectral type later than B9 are well described by our disk models and we
discuss in some detail the PAH characteristics one can derive from the existing
data.  Objects with strong radiation field (generally earlier than about B9) have the 3.3
$\mu$m feature (often the only one observed) much weaker than predicted, and we
discuss possible explanations (dissipation of the disk, photoevaporation or
modification of the PAH properties).  
\keywords{circumstellar matter - stars: pre-main sequence - dust, extinction - radiative transfer - infrared: ISM: lines and bands} }

\authorrunning{Habart et al.}
\titlerunning{PAHs in disks}

\maketitle

\section{Introduction}

Features of PAHs (Polycyclic Aromatic Hydrocarbons) are detected in a number of pre-main-sequence stars of intermediate mass with circumstellar disks, with
characteristics that differ from those in the interstellar medium \cite[see, e.g.,][]{vankerckhoven2002}. 
An origin of these features in the disk surface layers, which are directly exposed to the
ultraviolet (UV) stellar radiation, has been suggested by various authors \cite[see, e.g.,][]{meeus2001}.  
Recently, PAHs emission has been resolved spatially in HD 100546,
HD 97048 \cite[]{boekel03} and WL~16 \cite[]{moore98,ressler2003}, and shown to
have the size  typical of circumstellar disk emission. 

The presence of PAHs on disk surfaces is interesting for a number of
reasons. First of all, as we will see in the following, PAHs are good tracers of
the presence of very small particles mixed with the gas at large altitudes above
the disk midplane, both near the star and in the very outer disk.  Their
emission can tell us if and where very small particles survive settling and
coagulation processes that cause the majority of the original grain population
to grow to very large sizes  in the same objects
\cite[see, e.g.,][]{natta2003}.
A second point of interest is that the absence or
presence of PAHs will have a significant
impact on the gas physical properties and on the structure of the disk itself,
since in photo-dissociation regions we believe that they contribute a large
fraction of the gas heating \cite[via the photoelectric effect, see, e.g.,][]{weingartner01,habart2001a} and could dominate the H$_2$ formation on dust
surfaces \cite[]{habart2004}.

The presence and properties of PAHs in circumstellar disks can be tested by
comparing the observations
to the predictions of disk models, which have been shown to reproduce well the
global spectral energy distribution (SED) of the same stars
\cite[]{dullemond2001,dominik2003}.  We are interested in particular to check if
the disk models can account for the observed intensity of the various features
and their spatial distribution, when known.  To the best of our knowledge, there
are no such models available for disks around pre-main-sequence stars. These
disks are optically thick to the stellar radiation and thus require a proper
treatement of the radiation transfer. Recent models of the SED of HD~141569A (a
young B9.5 star with a complex debris disk) by \cite{li2003}, which include
PAHs, assume an optically thin disk, as appropriate for that particular object.

The paper is organized as follows. \S 2 provides a summary of the available
observations of PAHs in intermediate-mass stars. The disk models and the
radiation transfer scheme we use are described in \S 3.1 and 3.2.  The dust on
the disk surface is a mixture of large grains, which reach equilibrium with the
stellar radiation field, very small grains and PAHs, which are transiently
heated by the UV stellar radiation. The details of the adopted dust model are
given in \S \ref{model_dust}. The results of the calculations are described in
\S 4, which also explores the effect of varying the stellar, disk and PAHs
properties. In \S 5 we compare the model results to the observations. \S 6
summarizes our main results.

\section {Summary of the Observations}
\label{observations}

\normalsize

From ground-based and airborne observations \cite[]{brooke93,schutte90},
it is known that aromatic infrared emission bands (AIBs in the following)
attributed to PAHs exist in the spectra of some Herbig Ae/Be stars (hereafter HAeBe).
About 20\% of the HAeBe stars surveyed to date have firmly detected 3.3 $\mu$m
feature \cite[]{brooke93}.

In Table \ref{tab_obs_1}, we report for an ISO sample of HAeBe stars with
circumstellar disks (i) the astrophysical parameters of each star (distance,
spectral type, effective temperature, luminosity, and $\chi$, i.e., FUV flux at 150 AU from the
star), (ii) the observational characteristics of the disk (evidence for a flared
disk and disk outer radius) and (iii) the dust spectral characteristics
(presence of AIBs, silicate bands and crystalline signatures). We also give, when available, estimate of
the PAH emission extension.  The strength of the PAH emission features at 3.3,
6.2, 7.7, 8.6, 11.2(3), 12.7 and 16.4 $\mu$m (where most commonly observed PAH
features occur) are given in Table \ref{tab_obs_2}.  Most of these sources are
{\it isolated} HAeBe stars, with the exception of two objects, HD 97048 and
Elias 1, which are still embedded in a reflection nebula.  These sources are
intermediate between embedded and isolated HAeBe stars.  In Tables
\ref{tab_obs_1} and \ref{tab_obs_2}, we also report the characteristics of the
HAeBe stars observed at 3 $\mu$m by \cite{brooke93} and of the young stellar
object WL 16 embedded in the $\rho$ Ophiuchi cloud core \cite[for a recent paper see][]{ressler2003}.

In the ISO sample of isolated HAeBe stars, we can distinguish three different groups
\cite[]{meeus2001}: the first (group Ia) is made of stars with evidence of
flared disk and strong or moderately aromatic and silicate bands (with or
without crystalline signatures); the second (group Ib) is also characterised by
flared disks and presence of AIBs, but the silicate bands are absent; finally,
in the third group (group II), there is no evidence of flared disk, weak or no
AIBs, but the silicate emission features are strong (with various amount of
crystalline silicates).  The presence of AIBs is not correlated to any of the
stellar parameters. There is, on the other hand, an indication that PAH features
are strong when disks are flared, since they are present in group I objects
(both Ia and Ib) and only (and rather weakly) in one source of group II. This
suggests that PAH emission originates in the outer regions of flared disks,
where dust intercepts a large fraction of the UV emission coming from the star.

That in isolated HAeBe stars PAHs must be closely related to the star-disk
system, is supported by the observations of HD 100546 and HD 179128 (both group
Ia) showing that the PAH emission occurs within $\sim$150 AU
\cite[]{grady2001a,boekel03} and $\sim$600 AU \cite[]{siebenmorgen2000} from the
star, respectively.  Moreover, \cite{boekel03} find that HD 104237 (group II) is
unresolved in their 10 $\mu$m spectroscopic observations using TIMMI2 at the ESO
3.6-m telescope with a spatial resolution of $\sim$0.9''.  This is consistent
with the idea that group II sources do not possess a flared outer disk.  As far
as we know, no spatial information on the PAH emission is available for the
other isolated HAeBe stars but their isolated nature suggests that also in these
objects the PAH emission comes from the immediate circumstellar environment.  On
the other hand, for the two intermediate sources, HD 97048 and Elias 1, the PAH
emission probably originates in a significant part from the illuminating
reflection nebula.  HD 97048 observations show that the PAH emission is extended
up to $\sim$1000 AU \cite[]{siebenmorgen2000}.  However, recent high spatial
resolution observations show that most of the emission of the 8.6, 11.3 and 12.7
$\mu$m features comes in fact from a region of 200-300 AU, likely a disk
\cite[]{boekel03}.
Additional evidence of the PAH emission disk origin in HAeBe stars comes from
recent (sub)arcsec-resolution mid-IR images of WL 16 by
\cite{moore98,ressler2003}, showing that the PAHs emission has a size of
$\sim$450 AU.  Finally, note that for the HAeBe stars observed from the ground
at 3 $\mu$m, \cite{brooke93} finds that the aromatic features tend to be confined
to $\sim 1000$ AU.

It is also interesting that in HD 97048 and WL 16 the 10 $\mu$m continuum
emission is spatially extended on scales comparable to those of the PAH emission
\cite[]{boekel03,moore98,ressler2003}.  Together with the absence of
the spectral signature of silicate grains in both these sources (see references
in Table \ref{tab_obs_1}), this indicates that very small carbonaceous grains
are responsible for the continuum emission.  Quite differently, the continuum
emission in HD 100546 arises from a much smaller ($<$30 AU) region than the PAH
emission \cite[]{boekel03}.  In this object, where silicate bands are present,
the continuum may be dominated by warm large silicates grains confined to the
innermost regions of the disk.

\par\bigskip Evidence of a disk origin of the PAH emission in HAeBe stars comes
also from the extensive study of the ISO spectra of a sample of embedded young
stellar objects, isolated HAeBe stars, evolved stars, reflection nebulae and HII
regions, which reveals the presence of variations in the profiles and peak
positions of the main bands from source to source
\cite[]{hony2001,vankerckhoven2002,peeters2002}.  The largest variation is seen
in the 6-9 $\mu$m region with systematic differences 
between sources which have recently synthesized their PAHs in their ejecta 
and sources which are illuminating general interstellar medium materials \cite[]{peeters2002}.
Combined effect of PAH family and anharmonicity (dependence of the peak position
with the temperature of the emitters) could be at the origin of these
variations. The 6.2 and 7.7 $\mu$m features in isolated
HAeBe stars which resemble those in evolved stars are very different from those in sources with ISM material.
 Moreover, large variations in the ratio of the band strengths in
the 10-15 $\mu$m region are observed, especially in the 11.2/12.7 $\mu$m ratio
\cite[]{hony2001}. Together with the good correlation found between the 11.2 and
3.3 $\mu$m band and between the 12.7 and 6.2 $\mu$m band, \cite{hony2001} suggest
that the spectra of HII regions where the 11.2 and 12.7 $\mu$m are typically
equally strong are due to small or irregular (and preferentially ionized) PAHs.
On the other hand, the spectra of evolved stars with a small 12.7/11.2 $\mu$m
ratio may arise from large ($\sim$100-150 C-atom) compact (and preferentially
neutral) PAHs.  The 12.7 $\mu$m band cannot be seen in the low resolution spectra
of isolated HAeBe stars and the 12.7/11.2 $\mu$m ratio is not known.  But recent
ground-based observations \cite[see reference in][]{vankerckhoven2002}
confirm that, if present, the 12.7 $\mu$m band must be very weak compared to the
11.2 $\mu$m. This indicates that the PAHs are more large and less ionized in
isolated HAeBe sources.

In summary, there are indications that PAHs in HAeBe stars are somewhat
different from those in the ISM, and more similar to PAHs in
evolved stars.  However, it is difficult to fully characterize
the PAHs properties in HAeBe stars from what is 
known at present, and in the following we will start our analysis from
the assumption that PAHs in the ISM represent a good first approximation
to the properties of PAHs in disks as well. We will then discuss
possible differences, and come back to this point in our conclusions.

\section {Models}
\label{model}

In this section, we describe the disk models and the radiation transfer used and
give the details of the adopted dust model.

\subsection {Disk Models}

We consider disks heated by irradiation from the central star, in hydrostatic
equilibrium in the vertical direction, with gas and dust well mixed (flared
disks).  We compute the disk structure using an improved version of the
\cite{chiang97} two-layer models, described in \cite{dullemond2001}.  For a
given star, the disk structure (i.e., pressure scale height \Hp\ and flaring
angle $\alpha$) is completely defined once the inner and outer radii (\Ri\ and
\Rd), the surface density distribution $\Sigma=\Sigma_0 (R/R_0)^{-p}$ with $R_0$
a fiducial radius, and the dust model are specified.

In these models, the stellar flux impinging with the flaring
angle $\alpha$ into the disk
is absorbed in the upper layers of the disk, which will reradiate half of
the flux away from the disk and half down into the disk's deeper layers. 
Here, we compute the structure of the disk self-consistently using
an iterative scheme that gives the flaring angle $\alpha$ and the midplane
temperature at each radius.  The radiation transfer is solved in an effective,
simplified way, which, however, gives a disk structure very similar to that
obtained with a full, much more time consuming radiation transfer treatement
\cite[]{dullemond2003}.

The models are appropriate for disks that are optically thick to the stellar
radiation. This is generally the case for disks around pre-main--sequence stars,
up to very large radii (e.g., 5000 AU for a disk mass \Md$\sim 0.2$ \Msun,
$p=1$).

\subsection {Radiative Transfer}

In spite of the success of the two-layer codes in predicting the disk structure
and the overall SED, they tend to overestimate the disk emission in the
near-infrared \cite[]{dullemond2003}.  We have therefore decided to implement a
full 1D radiation transfer scheme to compute more accurately the emission in the
wavelength range where the PAH features occurr.  In practice, we first compute
the disk structure using our two-layer code, with the adopted dust model (which
includes transiently heated species), then, keeping fixed the disk structure, we
compute the emerging spectrum using the radiation transfer code.  This procedure
is not fully self-consistent, but we have checked that it does not introduce any
significant inaccuracy in our results.

Dust scattering is incorporated.  Scattering of stellar light may be relevant
because it can lead to a substantial reduction of the flux available for heating
the disk and because it allows stellar photons, after one scattering event, to
penetrate much deeper into the disk then one would expect given the smallness of
the flaring angle $\alpha$.

To calculate the emission of transiently heated small grains and of PAHs, we
compute the temperature distribution functions $dP/dT$ as in
\cite{siebenmorgen92} following the method outlined by \cite{guhathakurta89}.

To determine the emission from the disk, we divide it into rings of radius $r$
and small radial width $\Delta r$.  At the center, at $r=0$, sits the star with
parameters \Mstar, \Lstar, \Rstar\ and \Tstar.  For each ring, we compute, as
described in the appendix, the radiative transfer perpendicular to the disk, in
$z$--direction, and then add up the contributions from all rings.

\subsection{Dust model and PAH properties} 
\label{model_dust}

The properties of dust in circumstellar disks differ significantly from those in
the ISM (see \S \ref{observations}), and from star to star. It is difficult to
build a general, realistic dust model. However, this is not the purpose of this
study, where we only want to test the hypothesis that the observed PAH features
have their origin in the surface layers of circumstellar disks.  Therefore, we
will start with a dust model that includes large grains (i.e., grains large
enough to be in thermal equilibrium with the local radiation field) with
properties that roughly account for the observed SEDs (including the silicate
emission at 10 $\mu$m) of most pre-main-sequence stars. 
To these, we will add very small particles and PAHs, with abundances and properties
typical of the ISM, that we will then vary.

\par\bigskip\noindent {\it a) Large grains.} The large or big grains (BGs
in the following) consist of graphite and silicates with optical constants from
\cite{draine85}. These grains have a MRN size distribution ($n(a) \propto
a^{-3.5}$) between a minimum and a maximum radius $a_l=$0.01 $\mu$m and
$a_u=$0.74 $\mu$m for silicates and $a_l=$0.01 $\mu$m and $a_u=$0.36 $\mu$m for
graphite.
We assume that
all silicon and about 2/3 of the cosmic carbon\footnote{\cite{snow95} have
discussed the carbon abundance in stars newly formed from the ISM and conclude
that the carbon abundance in the local ISM is (2.24$\pm$0.5)$\times 10^{-4}$.}
($[C/H]_{BGs}=1.5\times 10^{-4}$) are in these grains. This prescription gives a
good fit to the mid-IR SED (which originates from the disk surface) of many HAe
stars \cite[]{natta2001}; we have found in our modeling that the details are not
very important for the purpose of this paper.

\par\bigskip\noindent {\it b) Very small grains.} To the BGs population of
grains, we add a tail of very small graphite particles (VSGs in the following)
with the same size distribution $n(a) \propto a^{-3.5}$ between $a_l=$10 \AA\ to
$a_u=$100 \AA.  Optical constants are from \cite{draine85}.  The fraction of
carbon in VSGs is 15\% that in BGs ($[C/H]_{VSGs}=0.22\times 10^{-4}$).  VSGs
and BGs graphite grains probably form a continous population.  The division
between BGs and VSGs is maintained mainly for computational purposes.  Note that
there is no evidence in the ISM of very small silicates \cite[]{li2001b}.

\par\bigskip\noindent {\it c) PAHs.} Finally, we include in our dust model
a population of PAHs.  In the ISM, PAHs are made up of a few tens up to a few
hundreds carbon atoms; for reasons of simplicity, we start our models with one
PAH size, $N_C=100$.  The hydrogen to carbon ratio is $H/C = f_H \times
(6/N_C)^{0.5}$ \cite[case of compact symetric PAHs, see][]{omont86} 
with $f_H$ the hydrogenation fraction of the molecule. 
We start our models assuming essentially fully 
hydrogenated PAHs, i.e, $f_H=1$. 
The carbon locked up in PAHs has an abundance of $[C/H]_{PAH} \simeq 5 \times 10^{-5}$ inferred from the 12 $\mu$m emission per hydrogen in typical Galactic
cirrus \cite[]{boulanger88} and photo-dissociation regions \cite[]{habart2001a}
and from comparison between observations of dust galactic emission and
extinction with detailed model calculations \cite[]{desert90,dwek97,li2001a}.
We take the absorption cross section as defined by \cite{li2001a} based both on
laboratory data and astrophysical spectra.  They consider the bands at 3.3, 8.6,
11.3, 11.9 and 12.7 $\mu$m from vibrational modes of the aromatic C-H bond; the
strong bands at 6.2 and 7.7 $\mu$m due to vibrations of the aromatic C-C bonds;
and a few weak features probably caused by C-C bending modes at 16.4, 18.3, 21.2
and 23.1 $\mu$m.  The band profiles have a Drude shape; the integrated cross
sections band profiles, positions and widths of the lines are found in their
Table 1 \footnote{In order to fit the observed IR spectrum of the diffuse ISM,
\cite{li2001a} include enhancement factors $E_{6.2}$, $E_{7.7}$ and $E_{8.6}$ in
the integrated cross sections for the 6.2, 7.7 and 8.6 $\mu$m bands.  For the
template model, we adopt $E_{6.2}$, $E_{7.7}$, $E_{8.6}$ equal to 1, which
recover typical laboratory values, but we will also consider model with
$E_{6.2}$, $E_{7.7}$, $E_{8.6}$$>$1 (see \S \ref{cross_section}).}.  The cross
section in the visible-UV is determined from the laboratory absorption spectrum
of small species with a cut-off in the visible to near-IR whose wavelength
increases with $N_C$, as defined by \cite{desert90}.

With respect to the charge, we start assuming that PAHs are mostly neutral.
Ideally, one can calculate the steady state charge distribution of PAH from the
balance between electron capture rates and the photoelectron emission rates plus
the ion capture rates \cite[see, e.g.,][]{weingartner01}.  Since little is known
regarding the electron density $n_e$ and its distribtion in the surface layers
of disks, precise determination of the PAH charge distribution in disks is not
possible.  Instead, we can estimate the ratio between the photoionisation rate
of the grain to the electron-grain recombination rate, $\gamma$, which can be
approximated by $\gamma _0 \times f(N_C) \ N_C^{2/3} \times \chi
~T_{gas}^{1/2}/n_e$ with $f(N_C)$ a function giving the dependence of the
photo-electron ejection probability with the size \cite[see][]{bakes94}.  For a
disk heated by a typical HAe star, we have $\chi \sim 10^5$ and $n_H \sim 10^7$
cm$^{-3}$ on the surface disk layer at 150 AU from the star (see next section).
Thus, assuming $n_e/n_H \sim 1.5~10^{-4}$ (as in typical photo-dissociation regions of the ISM)
 and $T_{gas}\sim 100$ K, we find $\chi ~T_{gas}^{1/2}/n_e \sim 500$ and
for a PAHs with $N_C\sim 100$ we estimate\footnote{To estimate $\gamma$, we take
$\gamma _0 \sim 3 \times 10^{-6}$ and $f(N_C)\sim 11$ as in \cite{bakes94}.}
that $\gamma$ is inferior than 1.
This implies that PAHs must be neutral or negatively charged in the outer disk regions.
In the inner disk region and for stars with higher $\chi$, however, $\gamma$
becomes higher and PAHs will be in part positively charged.  To keep the model
simple, we neglect that we probably have a mixture of neutral and charged PAHs
in disks and we assume that PAHs are neutral.  The effect of charge is briefly
discussed in \S \ref{charge}.

Upon absorption of an energetic photon, small PAHs, with an insufficient number
of internal vibrational modes in which to distribute this photon energy may
dissociate.  In a strong FUV radiation field ($\chi > 10^2$), destruction occurs
during multiphoton events where PAHs absorb an energy of more than 21 eV in a
time shorter than the cooling time \cite[]{guhathakurta89,siebenmorgen93}.
Recently, \cite{li2003} estimated the photodestruction rate for the HD 141569A
disk
and suggest that PAHs with $N_C \le 40-50$ will be photolytically unstable in
the inner $R<100$ AU region around the star (where $\chi > 10^5$) during the
life time of the disk.
For a PAH with $N_C\sim 40-50$, the maximum temperature reached just after the
deposit of an energy of 21 eV is about 2000 K \cite[see, e.g.,][]{omont86}.  In
our model, we assume that PAHs are destroyed when the probability that they are
at a temperature $T\ge T_{evap}$ exceeds a critical value $p_{evap}$, i.e., when
$\int^{\infty}_{T_{evap}} P(T) dT \ge p_{evap}$.  Such procedure is numerically
simple and can probably correctly evaluate evaporation.  However, we do not
derive $p_{evap}$ from thermodynamical considerations.  We use $T_{evap}=2000$ K
and $p_{evap}=10^{-8}$ and checked that for this choice, in the case of small
PAHs, the results are similar to the assumption that the grains are
photo--destructed if they absorb an energy of more than 21 eV in a time interval
shorter than the cooling time.

\begin{figure}
\begin{center}
\leavevmode
\centerline{ \psfig{file=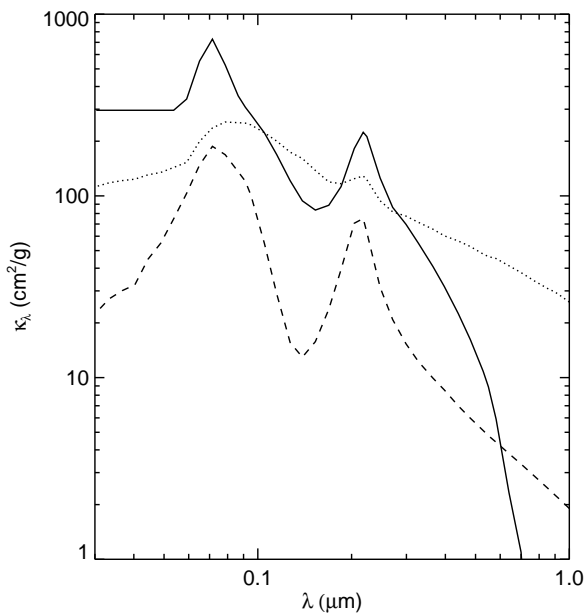,angle=0,width=9cm}}
\end{center}
\caption {Absorption coefficient (in cm$^2$ per gram of gas)
of large grains (BGs; dotted line), very small grains (VSGs; dashed line) and PAHs of 100 C atoms  (solid line). See text for details.}
\label{cross_sections}
\end{figure}

\par\bigskip In summary, we adopt a dust model with BGs, VSGs and PAHs.  The
silicate abundance in dust is $[Si/H]=3\times 10^{-5}$, and the total carbon
abundance in dust is $[C/H]=2.22\times 10^{-4}$.  Of this, 23\% is in PAHs, 10\% in
VSGs and 67\% in large grains.  Fig.~\ref{cross_sections} shows the adopted
absorption coefficient for the various species. For a radiation field that has a
blackbody distribution at T=10$^4$ K, BGs will absorb 50\% of the radiation, VSG
11\% and PAHs 39\%, respectively. The properties of BGs are somewhat different
from those of grains in the ISM, and have been chosen to approximately fit the
properties of dust on disk surfaces. VSGs and PAHs have properties derived from
the ISM. In the following, however, we will explore the effects on the predicted
spectra of some of our assumptions for the PAHs properties.  Note, however, that
we will not try to reproduce the peculiarities of the features observed in some
isolated HAe stars, such as the shift of the peak positions, but we will limit
our discussion to the band strengths.

\section{ Results}
\label{results}

In this section, we describe the results of the calculations and explore the
effect of various stellar, disk and PAHs properties.

\subsection{Template Model}
\label{template}

Firstly, we discuss the results for a model that we will use in the following as
a template.

The disk is heated by a typical HAe star with effective temperature \Teff=10\,500
K, luminosity \Lstar=32.0 \Lsun and mass \Mstar=2.4 \Msun.  At a distance of 150
AU, the FUV radiation field integrated between 912 and 2050 \AA\ is $\chi=10^5$
times the average interstellar radiation field \cite[$1.6\times 10^{-6}$ \Wm;][]{habing68}. 

\begin{figure}[h]
\begin{center}
\leavevmode
\centerline{ \psfig{file=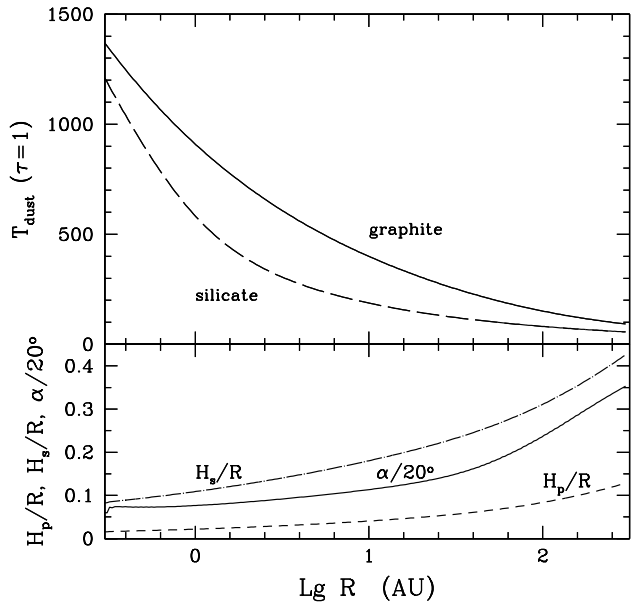,angle=0,width=9cm}}
\end{center}
\caption {Disk properties for the template model. The lower panel shows the run
with radius of the pressure scale height H$_p$, the photospheric scale height
H$_s$ and the flare angle $\alpha$ divided by 20 deg for convenience. $\alpha$ is 1.4 deg at the inner radius, and reached 7 deg at the outer disk radius.
The top panel shows the temperature for graphite (solid line) and silicate 
(dashed line) grains of size
0.01 $\mu$m at H$_s$.}
\label{disk}
\end{figure}

The disk mass is $\sim 0.1$ \Msun. Its inner radius is \Ri = 0.3 AU, close to
the BG evaporation radius, the outer radius is \Rd=300 AU. The surface density
profile is $\Sigma=\Sigma _0 \times (r/R_i)^{-1}$ with $\Sigma _0$= 2 10$^3$ g cm$^{-2}$.
We assume that only half of the star is visible from any point on the disk
surface ($\phi_\star=0.5$).  Fig.~\ref{disk} shows the run with radius of the
pressure (\Hp) and photospheric (\Hs) scale height and of the flaring angle
$\alpha$ (bottom panel).  In the top panel we show the temperature of silicates
and graphite grains of size $a=0.01$ \um\ at \Hs.  The gas density at \Hs\
decreases from $7\times 10^{9}$ \cmc\ at \Ri\ to $3\times 10^6$ \cmc\ at \Rd.

Each side of the disk intercepts and reprocesses about 10\% of the stellar
luminosity. The fractional contribution to the reprocessed radiation of BGs,
VSGs and PAHs are, respectively, 60\%, 13\% and 27\%.  Note that the fraction
emitted by PAHs is lower than the fraction of the stellar radiation they absorb
(about 39\%, see precedent section) because of their evaporation in the inner
part of the disk.  About 10\% of the total absorbed flux is lost in PAH
evaporation.

Fig. \ref{template} shows the calculated spectrum in the 2-40 $\mu$m range of 
a star/disk system seen face-on at a distance of 150 pc.
PAH features, which form in the optically thin disk surface, are clearly visible, and
some of them are very strong; in Table \ref{tab_mod_1}, we give the integrated
flux in the 3.3, 6.2, 7.7, 8.6, 11.3, 12.7 and 16.2 $\mu$m features after
continuum subtraction. 

The continuum is mostly due to the large grains which are very hot in the inner
regions (see Fig. \ref{disk}).  It is possible to see, for example, the broad
feature due to silicate emission peaking at about 10 \um\ under the much
narrower PAH features.  Some contribution to the continuum in the wavelength
range $\lambda <10$ \um\ is also due to the VSG, as shown by the SED of a model
which has all the parameters of the template model but no VSGs nor PAHs (dashed
line in Fig.~\ref{template}).

\begin{figure}[h!]
\centerline{ \psfig{file=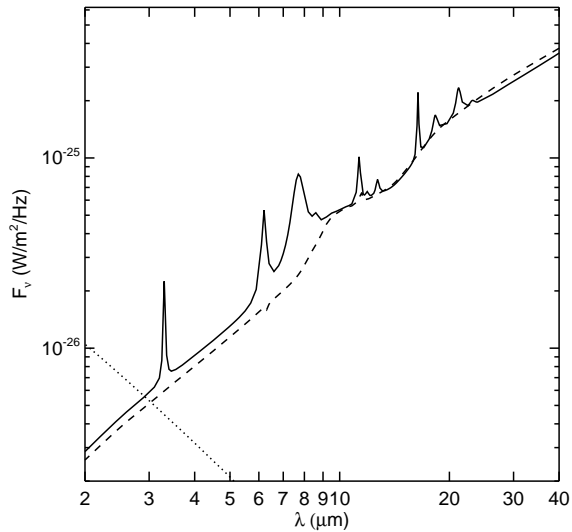,width=9cm,angle=0} }
\caption{\em Spectral energy distribution of a star/disk system
(template model) seen face-on at a
distance of 150 pc.  The solid line shows the emission of the disk, the
dotted line that of the stellar photosphere.
The dashed line shows the SED of a model which has all the
parameters of the template model but no VSGs and no PAHs.  The stellar
parameters are \Tstar=10,500 K, \Lstar=32 \Lsun, \Mstar=2.4 \Msun. The disk is
flared, it has a mass of 0.1 \Msun, an inner radius of 0.3 AU and an outer
radius of 300 AU (see Fig. \ref{disk}).}
\label{template}
\end{figure}

\begin{figure}[h!]
\centerline{ \psfig{file=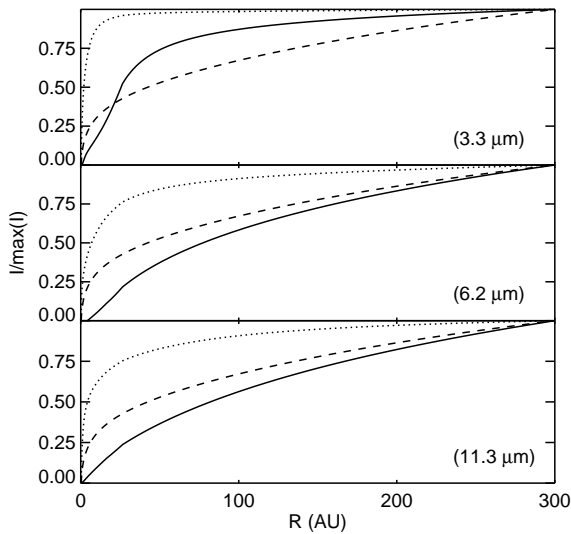,width=9cm,angle=0} }
\caption{\em Cumulative intensity of the feature (continuum subtracted, solid
lines) and continuum (dotted lines) at 3.3, 6.2 and 11.3 \um\ as
function of the projected radius.  We plot for comparison the run with radius of
the cumulative fraction of the FUV radiation intercepted by the disk surface (dashed lines).}
\label{template_cumint}
\end{figure}
\nopagebreak
\begin{figure}[h!]
\centerline{ \psfig{file=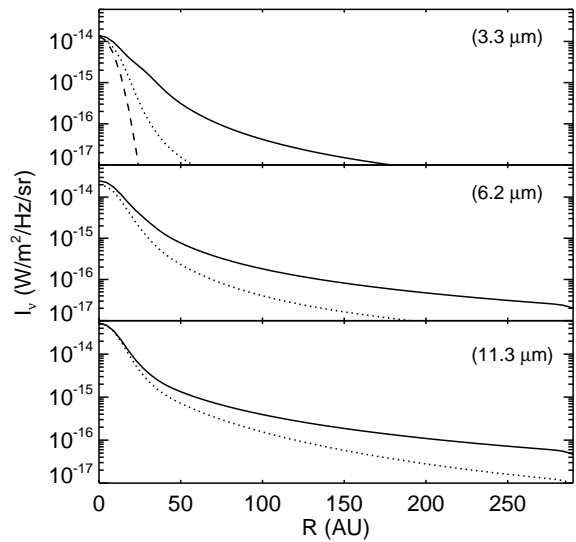,width=9cm,angle=0} }
\caption{\em Surface brightness profiles convolved with a two-dimensional
gaussian of FWHM=0.1'' at the peak of the feature (no continuum subtracted,
solid lines) and in the adjacent continuum (dotted lines) for the 3.3, 6.2 and
11.3 \um\ features.  In the top panel, we also show the
photospheric stellar emission at 3.3 $\mu$m convolved  with the same beam
(dashed line). As in Fig. \ref{template}, the star/disk is seen face-on at a
distance of 150 pc.}
\label{template_int}
\end{figure}

The PAH emission is much more extended than the adjacent continuum, as shown by
two figures, Figs. \ref{template_cumint} and \ref{template_int}.
The Fig.~\ref{template_cumint} plots as function of the projected radius the
cumulative intensity of the feature and continuum at 3.3, 6.2 and 11.3 $\mu$m. 
The continuum reaches 50\% of its intensity at a very small radius
(typically 2-5 AU) while the feature does so at larger radii (about 30 AU for
the 3.3 \um\ feature and 80 AU for the 6.2 and 11.3 \um\ features).  This
behaviour basically reflects the different excitation mechanism of PAHs, i.e.,
single-photon excitation, and of the continuum at the same wavelength, which is
dominated by the emission of BGs, in thermal equilibrium with the radiation
field. Only very hot BGs, located in the innermost disk, emit at 3 \um\ whereas
PAHs also emit at the outer edge of the disk in the 3.3 \um\ feature.

The ratio of short wavelength over long wavelength band strength decreases with
the intensity of the FUV radiation also for transiently heated particles.
Fig.~\ref{template_cumint} shows that the contribution of the outer disk to the
3.3 \um\ feature is small, while it is much larger for the features at 6.2 and
11.3 \um. We plot for comparison the run with radius of the cumulative fraction
of the FUV radiation intercepted by the disk surface. The similar behaviour with the
cumulative intensity of the less energetic PAH features reflects the fact that
the emission of PAHs roughly scales with the intensity of the FUV radiation
field (see \S \ref{star_parameters} for more details).

Fig. \ref{template_int} shows the intensity profile obtained by convolving the
computed intensity with a two-dimensional gaussian of FWHM=0.1''.  As expected
from Fig.~\ref{template_cumint}, the features at short wavelengths are strong in
the inner hot part of the disk decreasing rapidly in the outer cold part,
whereas those at long wavelengths are weaker in the inner region (mostly because
the continuum is higher)
but more extended (the temperature required to excite them is lower).
In other words, the 3.3 \um\ probes the inner ($R<100$ AU) region of the disk
while the other features probe its outer parts.

\subsection{PAH properties}
\label{pah_parameters}

In this section, we investigate the effect of the charge, hydrogenation
parameter, size, photoevaporation and absorption cross section of PAHs on the predicted intensity of the features.

\subsubsection{Charge state}
\label{charge}

In Fig. \ref{fig_ionised}, we show the spectra and the brightness distribution 
for the same model  shown in Figs. \ref{template} and \ref{template_int} 
but for PAH cations.  The PAH
spectrum is a strong function of the PAH charge: 
the strength of the C-C stretching modes between 6.2 and 8.6 $\mu$m is much
stronger for ionized PAHs; the 7.7 $\mu$m C-C feature to the 11.3 $\mu$m C-H feature ratio
 is increased by a factor of $\sim$5 (see Table \ref{tab_mod_1}).
Further, whereas neutral PAHs
show a strong 3.3 $\mu$m C-H features, the band almost vanishes when the PAHs
are positively ionized.  

The IR properties of PAH anions closely resemble those of PAH cations except for
the very strong 3.3 $\mu$m enhancement in the anion \cite[]{szczepanski1995,hudgins2000}.  However, 
\cite{bauschlicher2000} predict that PAH
anions have band strengths intermediate between those of neutrals (strong 3.3
\um\ C-H stretching and 11.3, 11.9 and 12.7 \um\ out-of-plane C-H bending modes)
and PAH cations (strong 6.2 and 7.7 \um\ C-C stretching and 8.6 \um\ C-H
in-plane bending mode).

\subsubsection{Hydrogenation parameter}
\label{hydrogenation}

\begin{figure}[h!] 
\begin{minipage}[c]{9cm}
\centerline{ \psfig{file=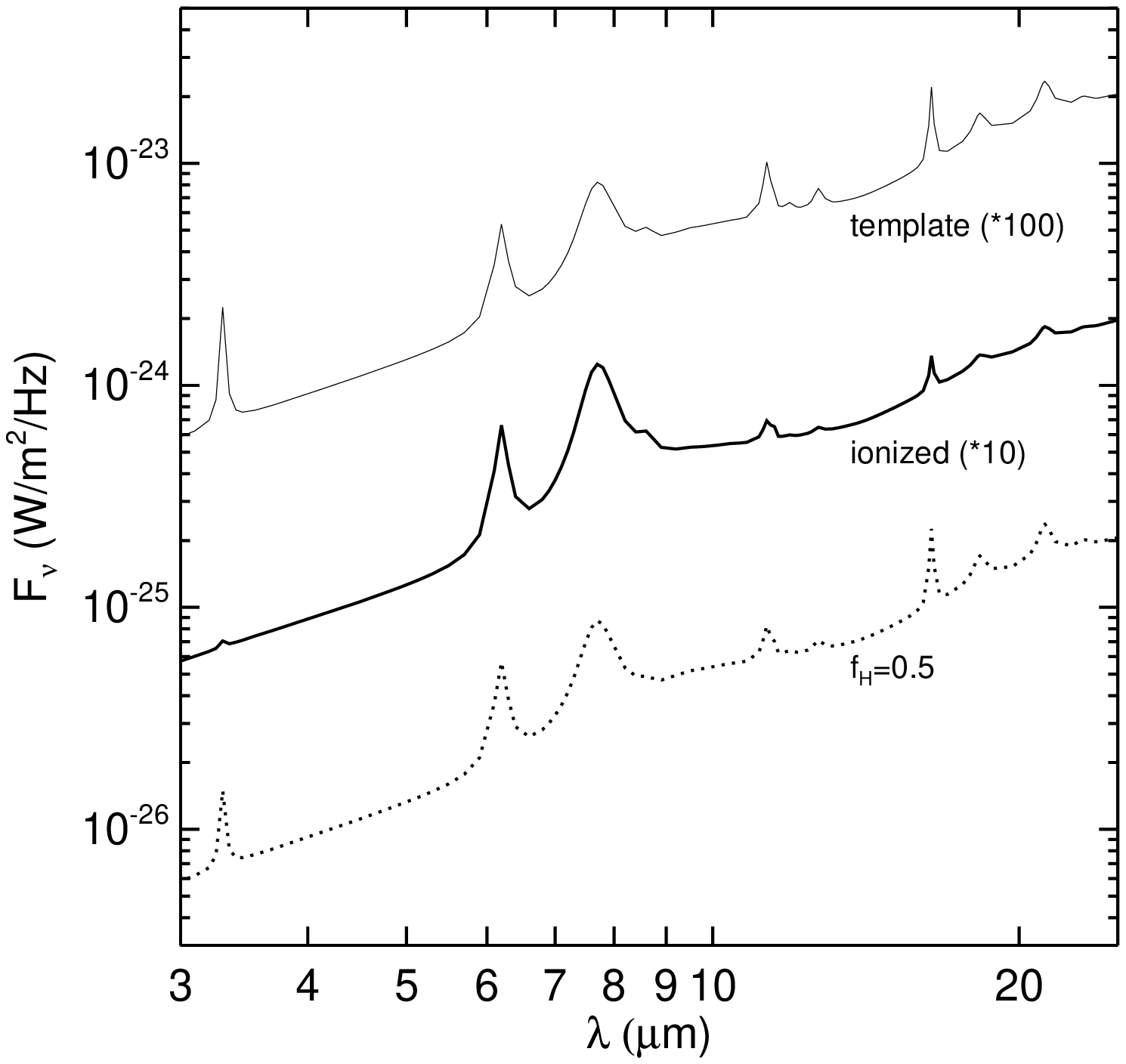,width=9cm,angle=0} }
\end{minipage}
\begin{minipage}[c]{9cm}
\centerline{ \psfig{file=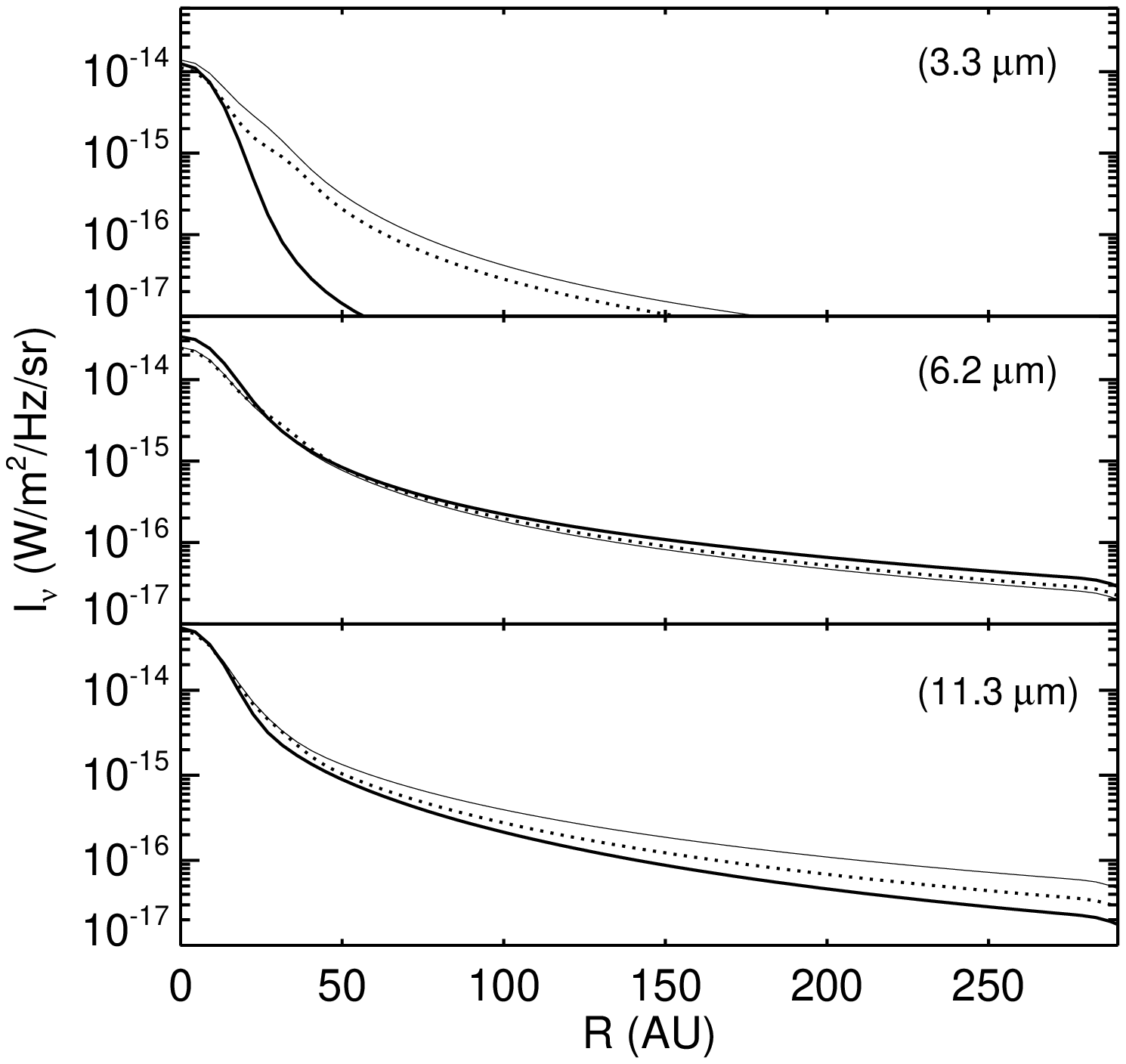,width=9cm,angle=0} }
\end{minipage}
\caption{ The top panel shows the SEDs of models
containing positively ionized PAH (solid thick line) and
partially dehydrogenated PAHs with $f_H=0.5$ (dotted line). 
All the other parameters are as in the template model. For
comparison, we show again also the SED of the
template model (solid thin line).
Note that the fluxes of the ionized PAH model and of the template model
have been multiplied by factors 10 and 100, respectively.
The bottom panel plots the brightness profiles convolved as in Fig. \ref{template_int}
at the peak of the 3.3, 6.2 and 11.3 $\mu$m features (no continuum
subtracted), as labelled. The lines refer to the ionized PAH model
(solid thick), the dehydrogenated PAH model (dotted) and to the
template model (solid thin).}
\label{fig_ionised}
\end{figure}

Another effect which can significantly affect the PAH spectra
is the dehydrogenation.    
Strong dehydrogenation would make the C-H features disappear.
However, theoretical studies of the dehydrogenation in the ISM 
indicate that dehydrogenation is relevant only for the
smallest PAHs with $N_C\le$25 \cite[]{tielens87b,allain96}.
Upon absorption of an energetic photon, large PAHs
will not have C-H bond rupture since the absorbed energy will promptly
be redistributed among many vibrational modes.
Nevertheless, in the inner disk regions and for stars with high
UV fields dehydrogenation could become important even
for large PAHs. To estimate this effect, we run a model with PAHs
partially dehydrogenated, i.e., hydrogenation fraction $f_H$=0.5.
The results are shown in Fig. \ref{fig_ionised}.  The C-H features (3.3,
8.6, 11.3, 12.7 $\mu$m) become lower (by a factor $\sim$2, see Table
\ref{tab_mod_1}), and the C-C features (6.2, 7.7 $\mu$m) become slightly higher
(as expected from energy conservation).

\begin{figure}[h!] 
\begin{minipage}[c]{9cm}
\centerline{ \psfig{file=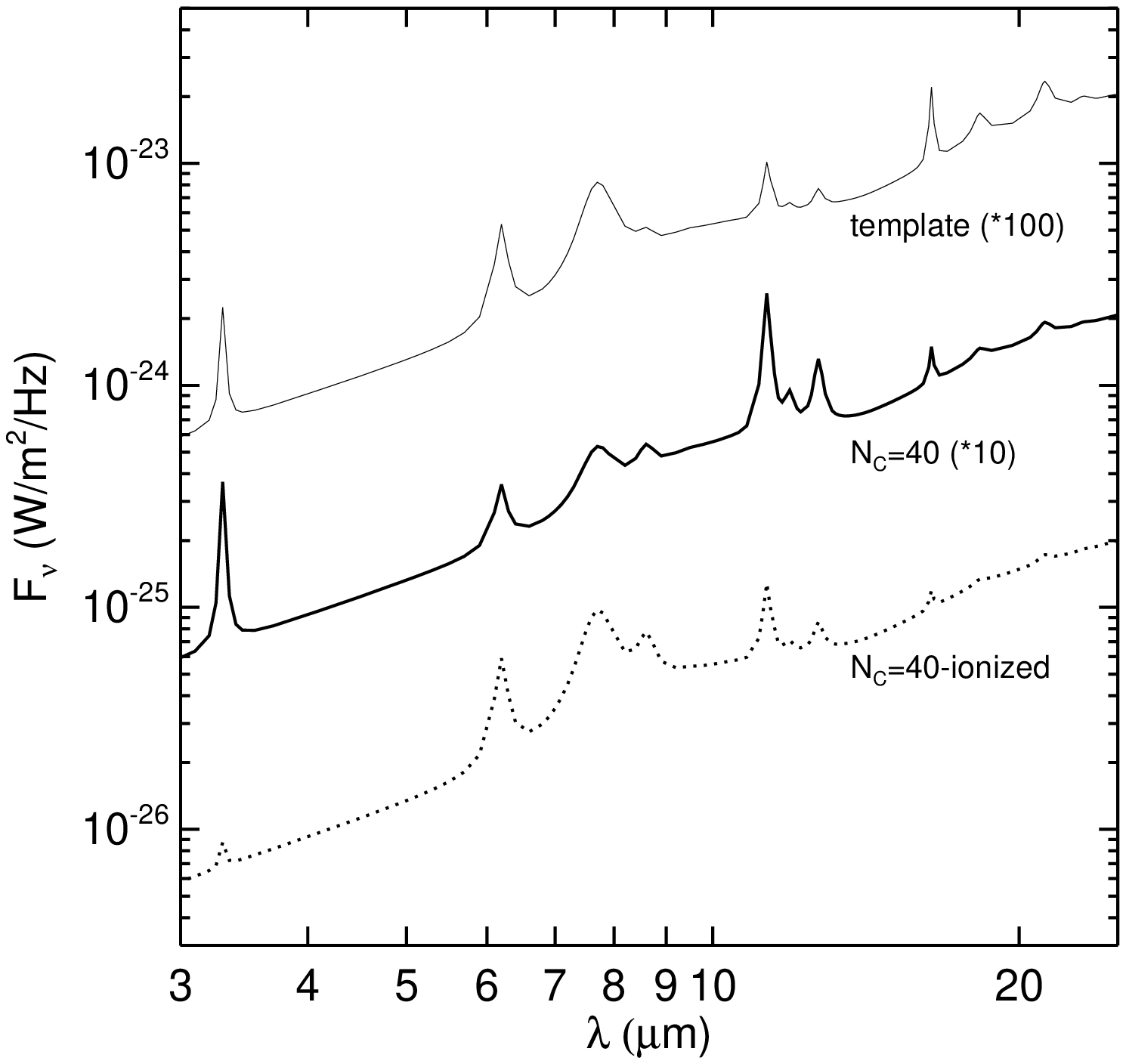,width=9cm,angle=0} }
\end{minipage}
\begin{minipage}[c]{9cm}
\centerline{ \psfig{file=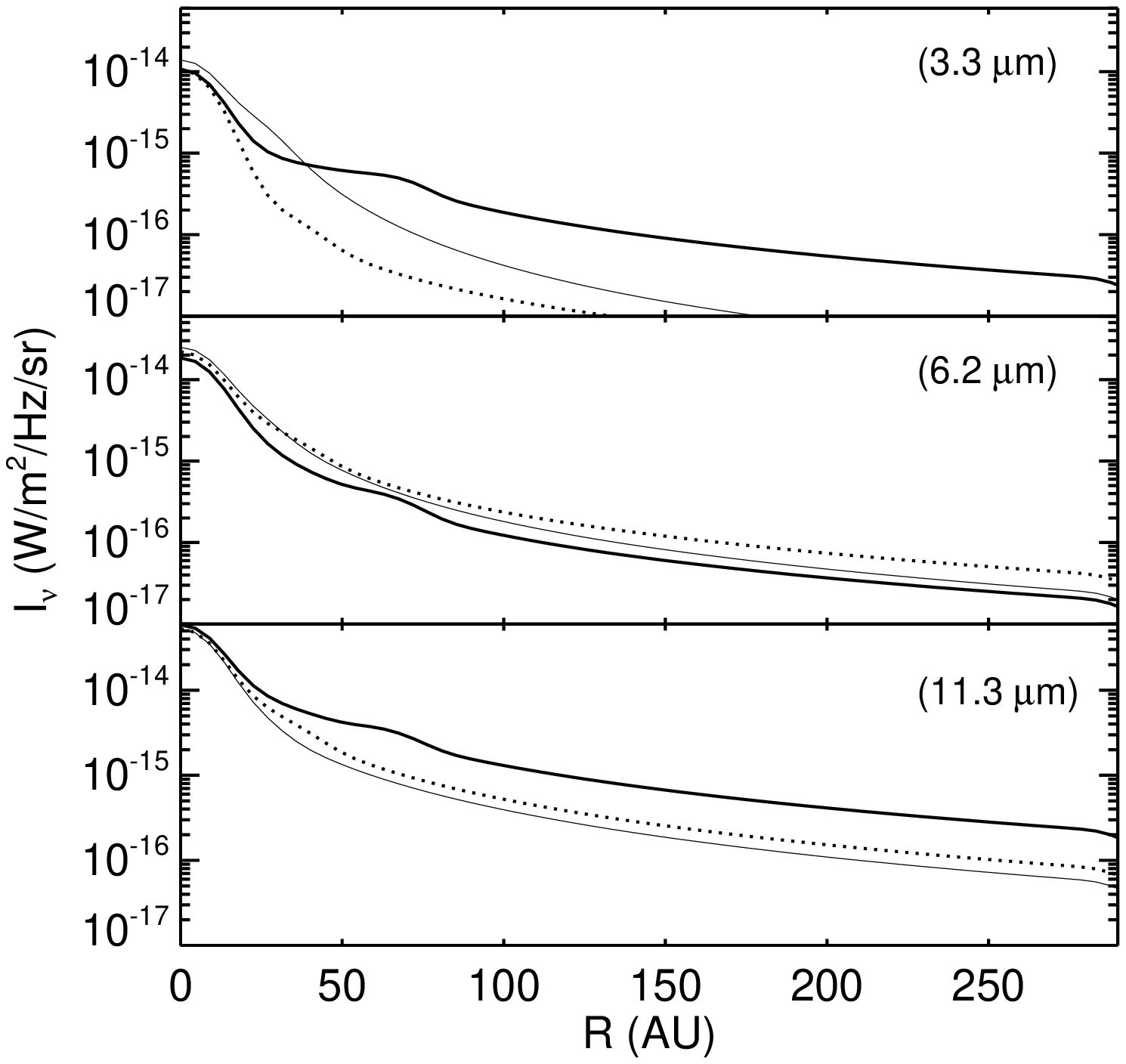,width=9cm,angle=0} }
\end{minipage}
\caption{\em Same as Fig. \ref{fig_ionised}  for model
containing neutral PAHs with $N_C$=40 (solid thick lines) and ionized PAHs with $N_C$=40 (dotted lines). Again, we show also the template model by solid thin lines.}
\label{fig_size}
\end{figure}

\subsubsection{Size}
\label{size}

We now look at the influence of the PAH size.
In Fig. \ref{fig_size}, we show a model for smaller PAHs with $N_C$=40. 
The relative strength of the PAH emission bands depends strongly
on the PAH size : small PAHs radiate strongly at 3.3 $\mu$m, while
larger (colder) PAHs emit most of their power at longer wavelengths.
The 3.3/6.2 $\mu$m ratio is $\sim 4$ times higher for small PAHs than for large PAHs (see Table \ref{tab_mod_1}).
In addition, since the H/C ratio decreases
with the size (see \S \ref{model_dust}), the C-H features increases
for small PAHs. 

We note that decreasing the PAH size goes generally in the opposite
direction than increasing the fraction of ionized PAHs:
the emission in the 6-9 $\mu$m range 
decreases for smaller PAHs while for ionized PAHs it increases (see \S \ref{charge}).
 On the other hand, the emission at 3.3 and 11.3 $\mu$m goes up when
PAHs become
smaller, whereas it goes down when the fraction of ionized PAHs increases.   

Therefore, we expect to find for small {\it and} ionized PAHs results similar to
those predicted by the template model with large and neutral PAHs.  For illustration,
we show in Fig. \ref{fig_size} a model with small  ($N_C=40$) {\it and} ionized
PAHs.  In this case, we find in fact that the strength of the 6.2, 7.7 and
11.3 $\mu$m features are roughly similar to those predicted by the template model (see
Table \ref{tab_mod_1}).  
However, size and ionization do not compensate each other fully and the 3.3 $\mu$m
band strength is considerably lower (by a factor of 10) than in the template
model.
But note that for small PAH anions, as mentioned in \S \ref{charge}, the 3.3 $\mu$m feature
strength will be higher.


Finally, we find that for small PAHs which are mostly destroyed in the inner
(hot) part of the disk -- $R\lesssim 70$ AU for $N_C$=40 -- the brightness
profiles of the features are affected.
The emission is decreased in the inner part and a plateau appears reflecting the compensation
between the graduated coming out of PAHs and the decrease of the FUV flux.
In the next section, we discuss the photoeveporation effect.

\subsubsection{Evaporation}
\label{evaporation}

With the formalism adopted here, we find that
 large PAHs with $N_C=100$ will not be photo-evaporated 
in most of the disk, except in the $R>20$ AU region, 
while smaller PAHs with $N_C$=40 will be mostly photo-evaporated in the $R\lesssim 70$ AU region 
(as mentionned above).
Larger PAHs are more stable since they can easily accomodate the absorbed photon energy.  
But PAHs evaporation, as discussed in \S \ref{model_dust}, is a
complex process, and it is possible that our treatment of it
underestimates its effects.
For stronger photoevaporation, the strength of the most energetic features which comes mostly from
the inner region will decrease and their brightness distribution will be affected;
the effect on the bands at longer $\lambda$ coming from
the outer region will be on the contrary negligible.

On the other hand, it is possible that continuous replenishment of PAHs via sublimation
of icy mantles (in which interstellar PAHs may have condensed during the dense
molecular cloud phase) or by accreting carbon atoms and/or ions from the gas
\cite[see, e.g.,][]{allain96} could maintain PAHs throughout all the disk. 
For the template model without evaporation, the 
strength of the 3.3 $\mu$m feature will increase by a factor of $\sim$2
while the other features at longer wavelength will not.

\subsubsection{Absorption cross section}
\label{cross_section}

Finally, we discuss the influence of the absorption cross section.
First, the PAH spectra depend on the integrated cross sections of individual
features, $\sigma ^{int}_{IR}$, which, unfortunately, differ among various lab
measurements or between experimental and theoretical studies.
To get an idea of how these uncertainties affect the PAH spectra we
 have computed models with enhancement factors of 2-3 in $\sigma
^{int}_{IR}$.
In particular, we have taken for the 6.2, 7.7 and 8.6 $\mu$m bands $E_{6.2}$=3, $E_{7.7}$=2 and
$E_{8.6}$=2 as suggested by \cite{li2001a} in order to fit the 6.2/7.7, 7.7/11.3 and
8.6/7.7 $\mu$m  band strength ratios observed in ISM spectra.
 Compared to the template model, the 6.2/7.7 and 7.7/11.3 $\mu$m ratios
are higher by a factor $\sim$2. 

Second, the intensities of the PAH bands depend on the absorption cross section in the
visible and UV which are rather uncertain.  The cross section ($\sigma ^{2D}_{UV}$) defined in \S
\ref{model_dust} has been derived for planar PAHs from
laboratory measurements of small species \cite[$N_C\le 30$,][]{joblin92} and, for
bigger species, from optical constants for graphite \cite[]{verstraete92}.  However, for astronomical PAHs with more complex 3-dimensional
structure very little is known.  To estimate the effect of this uncertainty we
have computed a model with $\sigma ^{3D}_{UV}$ for small graphite spheres as in
\cite{draine84}.  The power absorbed by $\sigma ^{3D}_{UV}$ is $\sim$2 times
lower than for $\sigma ^{2D}_{UV}$ and, consequently the intensity of the
features is reduced by $\sim$2.

\subsection {Star and Disk Parameters}
\label{star_disk_parameters}

The SED of the template model depends not only on the dust model, and in particular on
the PAHs properties, but also on the star and disk properties.  In this section,
we investigate the influence of star/disk system parameters on the PAH emission
features.

\subsubsection{Stellar radiation field}
\label{star_parameters}

\begin{figure}
\begin{center}
\leavevmode
\centerline{ \psfig{file=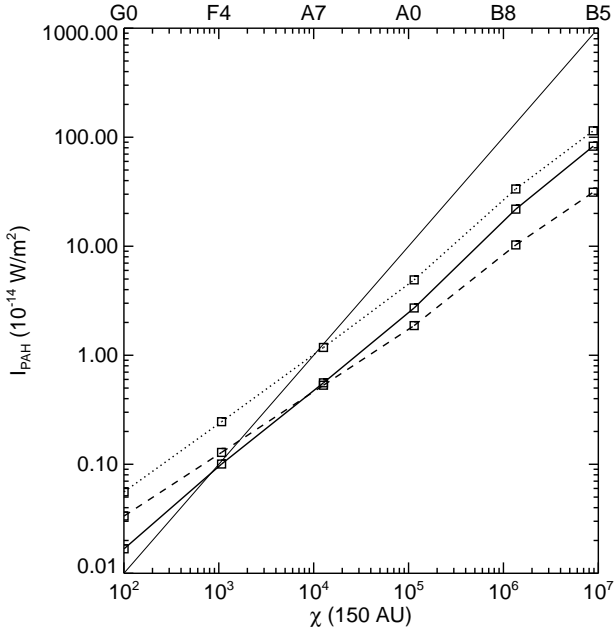,angle=0,width=9cm}}
\end{center}
\caption {Strength of the PAH features at 3.3 (solid line), 6.2 (dotted line)
and 11.3 $\mu$m (dashed line) for ZAMS stars.
The scale on the upper axys shows the spectral type,
that on the lower axys the corresponding value of
$\chi$.  The thin solid line shows a linear dependence I$_{PAH} \propto \chi$.
Disk and dust parameters are as in the template model,
which has  $\chi =10^5$
(\Lstar=32 \Lsun\ and $T_{\rm eff}$=10\,500 K).  }
\label{Int_pah_uv}
\end{figure}

The range of values of the FUV radiation field is large even within the rather
limited range of spectral types we investigate in this paper, with $\chi$ (the
FUV field at a distance of 150 AU from the star) varying from $10^6$ to few
$10^2$ (see Table \ref{tab_obs_1}).  $\chi$ changes because both the luminosity
and the effective temperature of the stars change. We have therefore computed a
set of disk models for stars of different spectral types assuming that they are
on the main sequence, and plotted the results in Fig.  \ref{Int_pah_uv} as
function of the corresponding value of $\chi$.  The spectral types are also
displayed in the figure.

As expected, the strength of the PAHs features increases with $\chi$.  The
dependence is not linear (see Fig. \ref{Int_pah_uv}),  because
the fraction of the stellar radiation absorbed by PAHs depends on its
wavelength dependence, i.e., on
$T_{\rm
eff}$.  For example,
the power absorbed by PAHs, normalised by $\chi$,
is two times higher for  $T_{\rm eff}$=10\,000 K than for 
$T_{\rm eff}$=15\,000 and ten times lower than for $T_{\rm eff}$=6\,000 K.

Fig. \ref{Int_pah_uv} shows the intensity of the  3.3, 6.2 and 11.3 $\mu$m
features as function of $\chi$. One sees that the 3.3 $\mu$m one increases
faster than the others, since it is the most sensitive to the
PAH excitation temperature. 
One can also note that the competing effect, i.e., the fact that PAH
evaporation is larger for higher $\chi$, reducing the 3.3 $\mu$m feature
more than the others, does not compensate entirely the effect of higher
excitation temperatures.
However, one should keep in mind that for smaller PAHs
($N_C< 100$), the effect of evaporation could become much larger.

\subsubsection{Disk geometry}
\label{disk_parameters}

The dependence of the PAH emission on disk parameters is easily understood.  The
quantity that most affects the resulting spectrum is the disk flaring angle,
which determines at each radius the fraction of FUV intercepted by the disk
surface. Lower values of the flaring can be caused by a variety of reasons, for
example if the dust settles toward the disk midplane. Also, if the disk mass is
fixed, a steeper dependence of the surface density profile than $p$=1 tends to
make the outer disk less flared, as the ratio of the photospheric to the
pressure scale height increases.  In all cases, the value of the flaring angle
in the inner disk is practically fixed, since it is dominated by its geometrical
part \cite[]{chiang97,dullemond2001}.  Therefore, less flared disks have lower
emission in all features, but a higher ratio of the 3.3 \um\ intensity, which
forms mostly near the star, where the flaring is dominated by the geometrical
part, to that of the 6.2 and 11.3 \um\ ones, which have a much larger
contribution from the outer disk.  In the extreme case of a geometrically flat
disk, the PAH feature strenghts decrease by 3 order of magnitude but the ratio 3.3/11.3 $\mu$m
increases by a factor of 3 and only the 3.3 $\mu$m band not completly disappears.
The 3.3 $\mu$m band to continuum ratio decreases by a factor
of 4 and becomes equal to 1.1.

In a similar way, when the disk is larger, all features get stronger, but
the 3.3 \um\ band less than the others.  If \Rd\ becomes very large, the disk turns
optically thin to the FUV radiation and the PAH emission does not increase further.

Another parameter to be considered is the disk inclination with
respect to the observer.
If the disk is inclined it will emit a lower continuum emission at long wavelength; 
for an inclination of $\theta =60$ deg ($\theta=0$ for face-on disks)
the models predict a continuum 
at $\lambda >$10 $\mu$m about 20\% lower than  face-on models,
while at shorter wavelengths there are practically no changes.
The continuum-substracted intensity of all the PAH features
we compute will not be affected at all.

A major source of uncertainty in our results derives from the lack of knowledge
of the detailed structure of the inner disk.  We have assumed in our
calculations that the disk is truncated at the dust sublimation radius, and that
each disk face is illuminated by 1/2 of the stellar surface.  In fact, there is
good evidence that in most HAe stars the disk has a puffed-up rim at the dust
sublimation radius, due to the direct illumination of the stellar radiation
\cite[]{natta2001,dullemond2001}.  The structure of this rim, and the effects of
the shade it projects further out on the disk are rather complex, depending on
the detailed 2D radiation transfer for a mixture of grains of different
properties.  The correct treatment of the effects of this region on the PAH
emission is well beyond the current capabilities of our models.  However, we can
make some qualitative estimates in the following way.  A puffed-up inner rim
will emit a strong continuum at short wavelengths.  This affects mainly the
ratio of the 3.3 $\mu$m band to the continuum; considering that models that
include the rim predict a 3.3 $\mu$m continuum a factor of $\sim 3$ larger than
our models \cite[see][]{natta2001}, the 3.3 $\mu$m peak to continuum ratio will
decrease by a factor of 2. However, the continuum-subtracted intensity of the
3.3 $\mu$m feature will not be affected; also, the effect on the other features
at longer wavelength is negligible.

The geometrical effect of the rim on the
disk illumination can be much smaller than we have assumed by taking
$\phi_\star=0.5$, to the limit of being entirely negligible.  In this case, each
point on the disk will be illuminated by the entire stellar surface, and the
fraction of intercepted stellar radiation will increase by a factor of 2 with
respect to our template model. We have checked that this results, with a large
degree of accuracy, in a factor of 2 increase of all the strong PAH features.
The total IR excess will also be a factor of 2 larger, going from about 20\%,
for a face-on disk, to about 40\%.  Most HAe stars have indeed IR excesses
within this interval \cite[]{meeus2001,dullemond2003}, and one expects that more
realistic models will be roughly intermediate between these two cases. However,
it should be noted that one can expect also some modest change in the ratio of
features at different wavelengths, as the rim shade will affect differently
different regions of the disk.

\section{Comparison with the observations}

\begin{figure*}[h!] 
\centerline{ \psfig{file=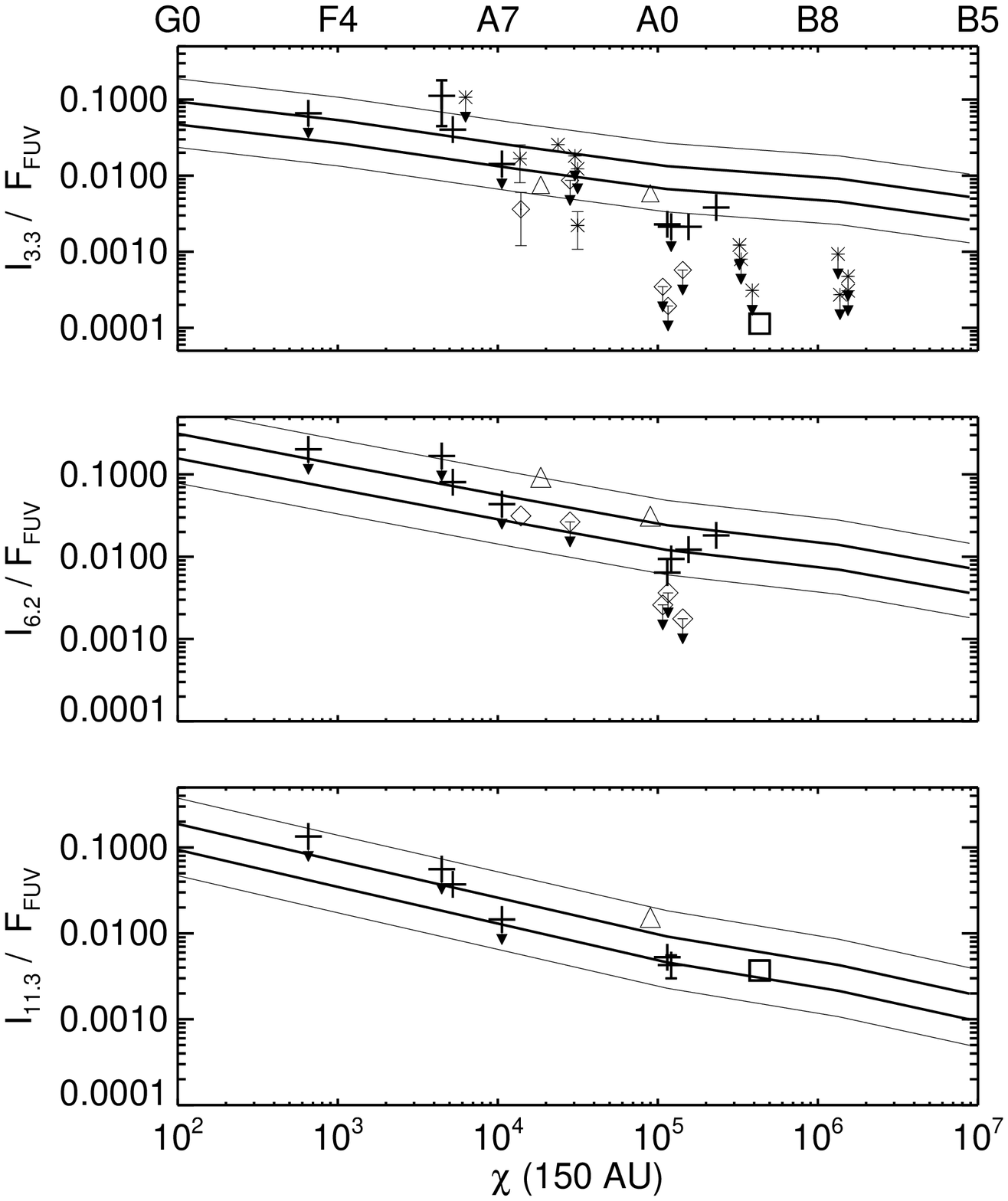,width=14cm,angle=0} }
\caption{Comparison of model predictions and
observations. \em {\bf Upper Panel} : Strength of the 3.3 $\mu$m feature
normalised to the FUV flux of the star for ZAMS stars. As in Fig. \ref{Int_pah_uv}, the 
scale on the upper axys is the spectral type, that on the lower axys
is the value of $\chi$, i.e.,
the FUV field at 150 AU from the star. The two solid lines
show the predictions of the template model for
$\phi_\star=0.5$ (lower curve) and $\phi_\star=1.0$ (upper curve)
(see text). The two thin lines show an uncertainty strip of $\pm$ a factor of 2
around these models.
The crosses show  observed intensities (normalized to the corresponding
FUV flux) for the isolated HAeBe stars with evidence of flared disk, the 
diamonds for the isolated HAeBe stars with no
evidence of flared disk and the triangles for 
the intermediate sources (HD 97048 and
Elias 1).  The asterisks show the 
HAeBe stars observed from the ground by
\cite{brooke93};  the square is WL 16,
corrected for dust attenuation (see text).  Error bars, when not shown, are
smaller than the symbols. Arrows are 5$\sigma$ upper limits.
{\bf Medium and lower Panels} : Same as upper pannel for the 6.2
and 11.3 $\mu$m feature.}
\label{comparison}
\end{figure*}

A first test of our models can be made by comparing the predicted intensity of
the most commonly observed PAH features (i.e., at 3.3, 6.2 and 11.3 \um ) to the
observations. We do that in Fig. \ref{comparison}, which plots for each feature
the intensity normalized to the FUV flux of the star, as function of the value
of $\chi$ (i.e., the FUV field at 150 AU from the star).  Both quantities are
distance-independent. The solid lines show the prediction of models for
main-sequence stars of different spectral type (and $\chi$), as described in \S
4.3.1.  The lower line refers to models where $\phi_\star=0.5$, the upper lines
to models with $\phi_\star=1$.  The thin lines show an uncertainty strip of
$\pm$ a factor of 2 around these models. This, as we have discussed, is a
conservative estimate of the uncertainties due to deviations from the
assumptions of the template models, in the disk description (different surface
density profiles, flaring etc.) and in the assumed dust model.

\subsection{PAHs in disks: consistent with observations?}

In the following, in order to see if PAHs in disks are consistent with observations, we check 
for each star group described in Sect. \ref{observations} if the disk models can account for the observed PAH  intensities.

\subsubsection{Flared disks}

Objects with flared disks, according to the classification of \cite{meeus2001}
(group I) are marked as crosses in Fig.~\ref{comparison}.  This group dominates
among detections, as expected from our models.  They are in general well
accounted for by disk models with ``standard" dust, as defined in \S
\ref{model_dust}.  This is true also for the two younger sources, i.e., HD 97048 and
Elias 1 (shown by triangles), especially if we correct for contamination in the
two longer wavelength bands by the associated reflection nebula \cite[about $\sim$40\% for HD 97048;][]{boekel03}.  We also show by asterisks the HAeBe stars
observed by \cite{brooke93} at 3.3 $\mu$m. They cover a range of $\chi$ from
about $10^4$ to very high values ($\sim 10^6$), and we do not have detailed
informations on their disk properties. However, the objects with $\chi\simless 10^4$ are consistent with our disk predictions.  No upper limit in objects
with $\chi \lesssim 10^5$ is significantly lower than the model predictions.

In 8 of these objects one has a measurement (or significant upper limit) of both
the 3.3 and the 6.2 $\mu$m features. The ratio ranges from 0.1 to about 0.5,
somewhat lower than the model predicted value of 0.5$\pm 0.2$. In four cases,
one can measure also the ratio of the 11.3 to the 6.2 $\mu$m feature. The values
are about 0.5 in 3 cases, and 0.8 (with an unceratinty of $\pm 0.2$) in the
forth object, in good agreement with the template model predictions of 0.4$\pm
0.2$.  It is likely, as we will discuss further in \S 5.3, that our models tend
to overestimate the intensity of the 3.3 $\mu$m feature by a factor 2--3 in stars
with $3 \times 10^4 \le \chi \le 3 \times 10^5$.
 
However, there are two groups of non-detections that are interesting. 
We discuss these ones in the following.

\subsubsection{Flat disks}

The first is the group of 5 diamonds, i.e., HAeBe stars with no evidence of flared
disk \cite[group II,][]{meeus2001}. Of these, only in 1 case (HD 142666) PAHs
have been detected at 3.3 and 6.2 $\mu$m; most of the non-detections are well
below the model predictions.  This agrees quite well with our result that flat
disks should have very weak PAH features, even when PAHs are present with normal
properties on the disk surface.

\subsubsection {High-$\chi$ stars}
\label{high_chi} 

Other interesting objects are those with high values of $\chi$ ($>3\times
10^5$). Most of them (the exception being WL 16, that we discuss in more details
in the following) are from \cite{brooke93}, and only the 3.3 $\mu$m feature has
been observed.  All these objects with high $\chi$ have 3.3 $\mu$m intensities
(or upper limits) well below the model predictions.

The interpretation of this group of objects is not straightforward; it is quite
possible that their disks, if they ever existed, have been dissipated by the
strong radiation field of the star \cite[]{natta2000,fuente2002}, or that they
are flat \cite[]{hillenbrand92}.
There are, however, other possibilites. One is that the abundance of PAHs is
lower than we assumed.  Photoevaporation, which increases rapidly with $\chi$
and for smaller PAHs, could easily reduce the PAH abundance well below our
assumed value.  PAHs evaporation, as discussed in \S \ref{model_dust}, is a
complex process, and we may have underestimated its effects.

It is also possible that the PAHs are ionized and/or dehydrogenated in the inner
disk region, resulting in a much reduced 3.3 $\mu$m band (see \S
\ref{hydrogenation} and \ref{charge}).  Observations of the PAH emission from WL
16, shown by a square in Fig. \ref{comparison} \footnote{In Fig. \ref{comparison}, the data for the embedded young
stellar object WL 16 has been corrected for dust attenuation taking a visual extinction of $A_V=31$
from \cite{ressler2003} and using the extinction coefficient of
\cite{draine84,draine89}.}, favors this last
possibility.  In fact, we find that while the models reproduce well the
11.3 $\mu$m feature strength, that of the 3.3 $\mu$m feature is well below the
predictions.  Moreover, \cite{ressler2003} interpret the spatial variation of
the 7.7 (or 8.6)/11.3 $\mu$m flux ratio as due to a change of the PAH charge
state along the disk.  In the inner region PAHs appear to be positively charged
while in the outer region they are neutral.  Furthermore, the 12.7/11.3 $\mu$m
flux ratio which increases in the outer part of the disk could indicate the
presence of more hydrogenated PAHs in the outer part.

\subsection{Which PAHs?}
\label{nature}

As we have just seen, our template model, with relatively large ($N_C=100$)
neutral PAHs fits very well the strength of the most commonly observed
features, with the possible exception of the 3.3 $\mu$m one, which tends to be
much lower than predicted in stars with $\chi \ge 3\times 10^5$.

In stars with lower $\chi$ (spectral type roughly later than B9), we can rule
out that PAHs are large {\it and} ionized, because they will produce
too low 3.3/6.2 and 11.3/6.2 $\mu$m ratios (by a factor of
$>$10 and 4, respectively, see \S \ref{charge} and Table \ref{tab_mod_1}). 
Similarly, large {\it and} strongly dehydrogenated PAHs are unlikely, since
they will also produce too low 3.3/6.2 and 11.3/6.2 $\mu$m ratios
(see \S \ref{hydrogenation}). 
Further, we can rule
out that PAHs are much smaller than our assumed model ($N_C=100$), 
because they will produce too high 3.3/6.2 and 11.3/6.2 $\mu$m ratios
(by a factor of 4 and 10, respectively, see \S \ref{size} and Table
\ref{tab_mod_1}). 

On the other hand, we cannot exclude that PAHs are small {\it and} ionized.  In
this case, we find that the strength of the 6.2 and 11.3 $\mu$m features, as well
as their ratio will be roughly similar to those predicted by the template model
with large and neutral PAHs (see \S \ref{size}).  This is easily understood
considering that the effects of the size and ionization compensate each
other. For small positively charged PAH, the compensation is nevertheless not
complete, and in fact, the intensity of the 3.3 $\mu$m feature will be lower than
in the template model by a factor $\sim$10 (see Table \ref{tab_mod_1}).
Therefore, this model, which underestimates the 3.3 $\mu$m feature strength by a
factor of 3 in stars with $3\times 10^4 \le \chi \le 3\times 10^5$ but by a much
larger factor in stars with $\chi \le 3\times 10^4$, seems unlikely.  For small
negatively charged PAHs, the 3.3 $\mu$m feature is expected to be stronger (see
\S \ref{charge}) and could become comparable to the observations.  However, it
is difficult to assess the reliability of this model, since the IR properties of
PAH anions are not well known.

To get some additional insight into the PAH properties, one can look at the
other weaker PAH features (i.e., 8.6, 12.7 or 16.4 $\mu$m) even if they have been
only sporadically detected. This is probably due to the low S/N ratio of the spectra.  
 The models with large neutral PAHs and small
ionized PAHs predict similar strengths of the 8.6 and 12.7 $\mu$m features, in
good agreement with observations (see Tables \ref{tab_obs_2} and
\ref{tab_mod_1}). On the other hand, large neutral PAHs will produce a 16.4 $\mu$m feature five
times stronger than small ionized PAHs.  This favors the small ionized PAHs
model since this feature has not been detected in the ISO spectra of isolated
HAeBe stars \cite[]{vankerckhoven2002}.  However, high S/N spectra in this
wavelength region are needed to determine the properties of the carrier of this
band.

The large and neutral PAHs hypothesis is supported by the detailed studies of
the PAH spectra in HAeBe stars based on both laboratory data and theoretical
calculations \cite[]{hony2001,vankerckhoven2002,peeters2002}.  Together with the
comparison of the PAH profiles observed in evolved stars and ISM-like sources,
they indicate that PAHs should be larger and less ionized in isolated
HAeBe stars, in agreement with our template model.

It must be emphasized that there are several complications which we have
neglected in this study. The most obvious is that we have assumed that PAHs can
be characterized by a single size and charge state.  This is unlikely to be the
case, and one can expect variations as a function of radius and depth in the
disk, as well as from object to object.  For example, PAHs are likley to be more
positively ionized in the inner disk region and for stars with higher $\chi$.
This could explain the gradual decrease of the 3.3 $\mu$m feature strength with
$\chi$ seen in Fig. \ref{comparison}.  In stars with $3 \times 10^4 \le \chi \le
3 \times 10^5$, it is possible that we have an {\it intermediate} situation,
and, in fact, a model with a mixture of both large neutral and small ionized
PAHs will predict a 3.3 $\mu$m feature strength lower than in the template model
by a factor of 2--3, close to the observed values.

We conclude that in stars with $\chi \le 10^5$ PAHs may be preferentially
relatively large ($N_C\ge$100) and neutral.  But small {\it and} negatively
ionized PAHs can also reproduce the observations.  For comparison,
\cite{li2003}, which have modelled the 
emission features from the (debris) disk around HD 141596A (with $\chi =10^5$)
conclude that PAHs must be mostly negatively charged.  However, they show that
models consisting of a mixture of both neutral and charged PAHs are also capable
of reproducing the observations.  For stars with $\chi > 10^5$, the large
neutral PAH model tends to overestimate the intensity of the 3.3 $\mu$m
feature. This could result from changes in the disk structure (e.g., disks may
have dissipated), or from changes in the characteristics of PAHs, which can be
ionized, dehydrogenated or strongly photoevaporated in the inner disk regions.

From the integrated spectra, it is difficult to discriminate between the various
effects due to ionization, dehydrogenation or evaporation.  In order to provide better 
insight into the PAH properties evolution within the disk, high angular spatial spectroscopic 
observations in the bands are needed. 

\section{Summary}

We have investigated the emission from PAHs on the surface of disks around HAeBe
stars by comparing model predictions with observations.  We have computed models
of disks, heated by irradiation from the central source, which contain
 large grains, in
thermal equilibrium with the radiation field, and transiently heated very small
grains and PAHs.  The disks are optically thick to the stellar radiation, in
hydrostatic equilibrium in the vertical direction, with dust and gas well mixed
(flared disks). We have used a 2-layers model to calculate the disk structure
and implement a 1D radiative transfer code 
to compute the emerging spectrum. This scheme is reliable
and efficient (see \S 3.1 and 3.2).  Our main results can be summarized as
follows.

\begin{enumerate}
\item 
The models predict an infrared SED showing PAH features at 3.3, 6.2, 7.7, and 11.3
\um\ clearly visible above the continuum.  The PAH emission, spatially extended,
comes mostly from the outer disk region ($R\sim$100 AU) while the continuum
emission at the same wavelength, mostly due to warm large grains, is confined to
the innermost disk regions ($R\sim$few AU).  Among the PAH features, the
3.3 $\mu$m one is the least extended; in our template model, about 1/2 of its
integrated intensity comes from $R<$30 AU.

\item 
From comparison with ISO and ground-based observations, we find that most of the
observed PAH features in objects with flared disks \cite[group I,][]{meeus2001}
are well described by our models.
In objects with no evidence of flared disk \cite[group II,][]{meeus2001}
no (or weak) PAH features have been detected, as expected from our models.
For geometrically flat disks, PAH features
are predicted to be very weak, even when PAHs are present with
standard properties on the disk surface.

%
 
\item Objects with strong radiation field (spectral type generally earlier than about B9)
 have the 3.3 $\mu$m feature (often the only one observed) much weaker than predicted.
We suggest that their disks have been dissipated by the strong radiation field, or
that PAHs have been photoevaporated more effectively than our model predicts, or
that their properties (ionization or dehydrogenation) have been modified in the
inner disk region.  All these scenarios will cause the 3.3 $\mu$m feature to
practically diseappear. 

\item Finally, we find that in circumstellar disks around stars with spectral type later
 than B9, PAHs should have an abundance typical of the ISM.
They may be preferentially large and neutral, or small and ionized,
and integrated spectra alone cannot distinguish between these two possibilities.
In order to provide better insight into the PAH properties evolution 
within the disk, high angular spatial spectroscopic observations
are needed. 
The fact that we can explain the observed PAH spectra, at least of the
isolated HAe stars, with PAH abundances and qualitative
properties similar to those
of PAHs in the ISM is interisting.
It seems that, if PAHs are reformed in disks,
after having been depleted (by coagulation or other processes) in dense
cores, they contain again roughly the same amount of carbon
(relative to that of large grains)
that they contain in the ISM.
This is a very interesting point, which certainly deserves further attention.

\end{enumerate}

To the best of our knowledge, this study represents the first successful
modeling of the PAH emission features from disks around pre-main-sequence stars.
It shows that PAHs are present in disks, at least around HAeBe stars,
and absorb a significant fraction of the stellar radiation, similarly to PAHs in the ISM.
Consequently, PAHs are an important source of opacity in circumstellar
disks and are likely to play a dominant role in the thermal 
budget and chemistry of the gas.  
The coupling between PAHs and gas can determine the
gas temperature in the outer disk regions, where the lower density can make the
collisional coupling between gas and grains inefficient, and as a consequence 
have a significant impact on the disk structure itself.
Finally, as a side aspect, PAH emission which is sensitive to the local
physical conditions - either directly or through a change of their physical characteristics -
could be useful to better understand the physical conditions prevailing in the surface of circumstellar disks.
Furthermore, PAH emission,
which is more extended than the thermal emission at similar wavelengths, can offer
an unique possibility to probe the outer disk structure.

\acknowledgements{We are grateful to F. Boulanger and L. Verstraete for fruitful discussions and their relevant comments and suggestions.}

\newpage
\normalsize
\par\bigskip\noindent
\par\bigskip\noindent
\par\bigskip\noindent
{\large Appendix: Numerical method for the radiation transfer}
\par\bigskip\noindent
\par\bigskip\noindent
 
The disk is divided into rings of radius $r$ and small radial width $\Delta
r$. At the center, at $r=0$, sits the star with parameters \Mstar, \Lstar,
\Rstar\ and \Tstar. For each ring, we compute, as described below, the radiative
transfer perpendicular to the disk, in $z$--direction, and then add up the
contributions from all rings.  \par\bigskip Each ring is treated as a
plane--parallel slab.  It displays mirror symmetry with respect to the mid plane
at $z=0$.  We solve the integral equations
\begin{equation}\label{RadTr_int1}
I^+(\tau) \ = \ 
e^{-\tau} \left( I^+(0) \ + \ \int_0^\tau S(x) \,e^x\, dx \right) 
\end{equation}
\begin{equation}\label{RadTr_int1a}
I^-(\tau)\ =\ e^{-\tau} \left( I_0 \ + \ \int_0^\tau S(x) \,e^x\,dx\right) \ 
\end{equation}

$I^+(\tau)$ is the intensity at optical depth $\tau$ and frequency $\nu$ in
upward direction along a straight line under an angle $\mu = \cos\theta$ with
respect to the $z$--axis; $\tau$ is zero at $z=0$.  The second equation
containing $I^-$ refers to downward beams.  $I_0$ is the intensity of radiation
that falls from outside (the star) on the disk. The indices $\theta$ and $\nu$
have been dropped for convenience of writing.  $S(\tau)$ is the source
function. The optical thickness $\tau$ and geometrical height $z$ are related
through $d\tau = \rho ~ \kappa ~ dz$ where $\rho$ is the mass density and
$\kappa$ the mass extinction coefficient of dust.

\par\bigskip There are two boundary conditions.  One states that at the top of
the slab the incoming intensity $I^- =I_0$ is zero except towards the star where
it equals $B_\nu$(\Tstar).  The stellar hemisphere, which seen from a distance
$r$ subtends a solid angle $\Omega = \pi R_{\star}^2/2 r^2$, is replaced by a
luminous band of the same intensity $B_\nu$(\Tstar). The band encircles the sky
at an elevation $\alpha$ and has a width $\Omega/2 \pi\ll \alpha$, its solid
angle is also $\Omega$.  The other boundary condition reads $I^+ =I^-$ at $z=0$.
\par\bigskip One starts the calculation of the radiative transfer for a ring
with a guess for the vertical dust temperature profile $T(z)$.  This fixes the
source function $S(\tau)$ once the dust properties have been specified.  Then
one solves (\ref{RadTr_int1}) and (\ref{RadTr_int1a}) for $I_\nu^+(z,\mu)$ and
$I_\nu^-(z,\mu)$ and calculates from $I^+, I^-$ the mean intensity of the
radiation field $J(z)$.  This allows to compute a new temperature run $T(z)$.
The procedure is iterated until $T(z)$ converges.

Because in our examples the visual opacity of the disk in $z$--direction can be
very large ($> 10^5$), we apply the following slight modification to what we
said above.  We slice the disk into a completely opaque mid layer sandwiched
between two thin top layers.  The mid layer, because of its high optical depth
and the lack of internal energy sources, is considered to be isothermal at
temperature $T_{\rm mid}$; the value depends, of course, on $r$.  We evaluate
the radiative transfer for the top layer according to the method described
before, and separately for the isothermal midlayer, but there it is trivial.  A
boundary condition connects the two layers.  In the calculation for the top
layer, we now position the zero point of the $z$--axis at the transition between
top and mid layer.  The intensity that enters the top layer from below under the
direction $\mu = \cos\theta$ is

\begin{equation}\label{stp_disk1}
I^+_\nu(z=0, \mu)\ = \ B_\nu(T_{\rm mid}) \Big(1-e^{-\tau_\nu(\mu)}\Big)\ 
\end{equation}

Here $\tau_\nu(\mu)$ is the optical depth of the mid layer.  The monochromatic
flux received by the top layer from below is therefore

\begin{equation}
F_\nu \ = \ 2\pi B_\nu(T_{\rm mid}) \int_0^1 \Big(1-e^{-\tau_\nu(\mu)} \Big)
\;\mu\,d\mu \
\end{equation}

The frequency integral $\int f_\nu \, d\nu$ must be equal to the downward flux
at $z=0$ because the net flux, for reasons of disk symmetry, vanishes
everywhere.  This condition yields $T_{\rm mid}$, of course, iteratively. In
mathematical form, it reads

\begin{equation}  
\int d\nu \, B_\nu(T_{\rm mid}) \int_0^1 \,d\mu \Big(1-e^{-\tau_\nu(\mu)} \Big)
\;\mu \ = \int d\nu \,  \int_0^1 d\mu \,\mu \, I_\nu^-(z=0)  \ 
\end{equation}

Ideally, the top layer extends downwards so deep that the temperature at its
bottom asymptotically approaches the temperature of the mid layer, $T_{\rm
mid}$.  Then the whole disk of arbitrary thickness is calculated consistently.
But that would require a visual optical depth of the top layer in
$z$--direction, $\tau_{\rm top}$, of several hundred and therefore many
(hundreds of) vertical grid points.  
In practice, we have found that $\tau_{\rm top} \simeq 4$ is sufficient, since
pratically all starlight is absorbed and $T(z=0)$ is very closed to $T_{\rm mid}$.
Typically, changing $\tau_{\rm top}$ from 4 to 40 changes the emerging SED by less than 
20\% at all wavelengths.

\par\bigskip 
We wish to make a few further points:  {\it
a)} The radiation that is absorbed in the top layer and then reemitted at
infrared wavelengths may leave the top layer, be again scattered or absorbed
there, or may enter the mid layer.  {\it b)} The net flux is always zero, but
the upward and downward integrated fluxes, $F^+$ and $F^-$, do depend on $z$, so
always $F^+ +F^- =0$, but $dF^+/dz \ne 0$.  {\it c)} Good agreement in the
results of our radiative transfer code with another one is demonstrated under
the internet address
http://www.mpa-garching.mpg.de/dullemon/radtrans/benchmarks.

\bibliographystyle{natbib}


\begin{table*}[h!] \caption{Parameters of Herbig Ae/Be candidates. Evidence for a flared disk and disk outer radius from spectral energy
distribution modeling and millimeter interferometry observations. Presence of AIBs
and silicate bands. Crystalline signatures.}
\label{tab_obs_1}
\begin{tabular}{lllllllllllll}
\hline
\hline
Object& d & Sp. Type & $T_{eff}$ & \Lstar & $\chi$$^a$ & Flared& $R_{disk}$$^b$ & AIBs & Sil. & Cryst. & Ref. \\  
       & $[pc]$ &  & $[K]$  & \Lsun  &    & disk & $[AU]$ & & & & \\  
\hline
AB Aur & 144 & B9/A0Ve &  9750 & 47& 10$^5$ & $\surd$ & 140 & $\surd$ & $\surd$  & - & (1)\\
HD 100546& 103 & B9Ve & 11000 & 36& 10$^5$& $\surd$ & 300(150$^{b_1}$) & $\surd$ & $\surd$  &  $\surd$ & (1) \\
HD 142527 & 200 & F7IIIe & 6250 &31& 4 10$^3$ & $\surd$ &          & $\surd$ & $\surd$  &$\surd$$^c$  & (1)  \\
\vspace{0.3cm}
HD 179218 & 240 & B9e & 10000 &80 & 2 10$^5$ & $\surd$ & ($<$600$^{b_2}$) & $\surd$ & $\surd$  &  $\surd$$^c$ & (1)  \\
HD 100453& 114 & A9Ve & 7500 & 9&  5 10$^3$  & $\surd$ &  & $\surd$ & -  & - & (1)  \\
HD 135344& 84 & F4Ve & 6750 & 3& 6 10$^2$    & $\surd$ &  & ? & -  &  - & (1)  \\
HD 139614& 157 & A7Ve & 8000 & 12& 10$^4$   & $\surd$ &  & - & -  &  -& (1)  \\
\vspace{0.3cm}
HD 169142 & 145 & A5Ve & 10500 & 32 &10$^5$& $\surd$ & 220 & $\surd$ & -  & -& (1)  \\
HD 104237 & 116 & A4Ve & 10500 & 40 & 10$^5$& no &  & - & $\surd$ & ? & (1)  \\
HD 142666 & 116 & A8Ve & 8500 & 11 &10$^4$ &  no &   & $\surd$ & $\surd$ & $\surd$$^c$ & (1)\\
HD 144432 & 200 & A9Ve & 8000 & 32 & 3 10$^4$&  no &   & - & $\surd$ & ? & (1)  \\
HD 150193 & 150 & A1Ve & 10000 & 40 & 10$^5$ & no &   & - & $\surd$ & $\surd$ & (1)   \\
HD 163296  & 122 & A3Ve & 10500 & 30 & 10$^5$& no &  360 & - & $\surd$  & $\surd$& (1) \\
\hline
HD 97048 &180  &B9-A0  &  10000    & 31 & 10$^5$ & & (300$^{b_1}$-900$^{b_2}$)   & $\surd$ &  - & - &  (2) \\
Elias 1 & 150 & A0-A6 & 8000 & 21 & 2 10$^4$ &  & 70 ($<$200$^{b_3}$) & $\surd$  & $\surd$ & ?$^d$  & (2) \\
\hline
Lk H$\alpha$ 25 &800  &A0 & 9800    & 9 & 2 10$^4$&  & ($<$1200)   & $\surd$ &  &  & (3)   \\ 
HK Ori &460           &A5  & 8300    &12  & 10$^4$&  & (70)   & $\surd$ & -  & & (3)    \\ 
HD 245185 &400  &A2  & 9100    & 17 & 3 10$^4$& &  (110)   & $\surd$ &  & & (3)    \\ 
Lk H$\alpha$ 198&600  & A5  & 8300    & 6 & 6 10$^3$& &  (900)   & - &$\surd$  &  & (3)   \\ 
BD+61$^{\circ}$154&650  &B8  &  11200   & 330 & 10$^6$& & (900)   & - &  & & (3)    \\ 
T Ori &460  &B9  &  10700   &  83& 3 10$^5$ & &  (88)   & - &  - &  & (3)   \\ 
V380 Ori &460  &B9  & 10700     &85  & 3 10$^5$& & (150)   & - & $\surd$  & & (3)    \\ 
HD 250550 &700  &B7  &  12300   &240  & 10$^6$ & & (410)   & - & -  &  & (3)   \\ 
Lk H$\alpha$ 208&1000  &B7  & 12300    &210  &10$^6$ & & (1600)   & - &  &  & (3)   \\ 
VV Ser &440  &B9  &  10700   &100  & 4 10$^5$& & (410)   & - & -  & & (3)    \\ 
WW Vul &700  &A3  &  8700   &20  & 3 10$^4$ & & (88)   & - &  &  & (3)   \\ 
BD+46$^{\circ}$3471 &1000  &A0  &9800     &508  & 10$^6$& &  (190)   & - &  & & (3)    \\
Lk H$\alpha$ 233 &880  &A5  &   8300  & 30 & 3 10$^4$& & (320)   & - &  $\surd$ &  & (3)   \\
\hline
WL 16 & 125 & B8-A7 & 9000 & 240 & 4 10$^5$ &  & (450) &$\surd$ & - & - & (4)\\  
\hline
\hline
\end{tabular}
 
$\surd$ : detection, ?: possible detection, -: no detection. \\
$^a$ Far-ultraviolet (FUV, 6$<h\nu<$13.6 eV) flux at 150 AU from the star expressed in units of the average interstellar radiation field, $1.6~10^{-6}$ W m$^{-2}$ \cite[]{habing68}. \\
$^b$ In parenthese we give estimate of the PAH emission extension. References: 
(b1) \cite{boekel03}; (b2) \cite{siebenmorgen2000}; (b3) \cite{brooke93}. \\
$^c$ Possible blend with the PAH 11.2(3) $\mu$m band. \\
$^d$ \cite{hanner94}.\\
References: (1) \cite{meeus2001} and references therein;
(2) \cite{vankerckhoven2002} and references therein; 
(3) for astrophysical parameters \cite{hillenbrand92} (except for BD+46$^{\circ}$3471 see \cite{berrilli92}
and for VV Ser see \cite{testi98}); for PAH emission extension and AIBs detection \cite{brooke93};
for silicate emission features detection \cite{berrilli92};
(4) \cite{ressler2003} and references therein. 
\end{table*}

\normalsize

\begin{table*}[!h] \caption{Integrated strength of the PAH emission features.}
\label{tab_obs_2}
\begin{tabular}{lllllllll}
\hline
\hline
\small

Object & 3.3       & 6.2    & ``7.7'' & 8.6 & 11.2(3) & 12.7 & 16.4 & Ref.\\
       & $[\mu m]$ & $[\mu m]$ & $[\mu m]$ & $[\mu m]$ & $[\mu m]$ & $[\mu m]$ & $[\mu m]$ \\

\hline
AB Aur    &  $<$1$^a$   & 4.4(0.3)   & 4.2(0.6) & 2(0.2)   &  2(0.6) &- &- & (1)\\
HD 100546 &  2.5(0.5)  & 14.3(0.4)  & 19.2(0.8)  & 4.6(0.5)   & -$^b$&- &-  & (1) \\
HD 142527 &  1(0.6)   &$<$1.5    &  -  & -&  $<$0.5&- &- & (1) \\
\vspace{0.3cm}
HD 179218 &  1.7(0.2)  &8.1(0.4)    &14.4(0.6) & 1.6(0.4)   & - &- &-  & (1)       \\
HD 100453 &  1.3(0.2)  & 2.6(0.5)   &-&-    & 1.2(0.3)&- &-  & (1)\\
HD 135344 &  $<$0.5         & $<$1.5         & ?& - & $<$1&- &- & (2)\\
HD 139614 &  $<$0.5         &$<$1.5          & - &- & $<$0.5&- &- & (2)\\
\vspace{0.3cm}
HD 169142 &  1(0.2)   & 2.8(0.3)   &  - &- &  2.3(0.5)&- &- & (1) \\
HD 104237&  $<$0.5         & $<$1.5          & - &- & -&- &- & (2)\\
HD 142666 &   0.3(0.2)  & 2.6(0.3)   & - &-  & - &- &- & (1)  \\
HD 144432&  $<$0.5         & $<$1.5          & - &- & -&- &- & (2)\\
HD 150193 & $<0.08$$^a$    & $<$1.5          & - &- & -&- &- & (2)\\
HD 163296 &$<$0.2$^a$    &  $<$1.5   &  -    &   -  &  -  &- &- & (2)  \\
\hline
HD 97048  &    1.3(0.3) &  6.9(0.8)  & 11.8(1.3) & 2.4(0.6)  &  3.4(0.4)& 1(0.2) &- & (2)  \\
Elias 1   &   0.5$^c$(0.1)  & 6.1(0.6)   &  9.4(0.1)  & 2.3(0.1) & -$^d$ &- &- & (2)   \\
\hline
Lk H$\alpha$ 25 & 0.076(0.006)   &           &  & &&&& (3) \\ 
HK Ori          &0.087(0.045)   &           &  & &&&& (3) \\ 
HD 245185       &0.035(0.018)  &           &  & &&&& (3) \\ 
Lk H$\alpha$ 198& $<$0.15 &           &  & & &&& (3)\\ 
BD+61$^{\circ}$154 & $<$0.09 &           &  & &&&& (3) \\ 
T Ori           &$<$0.15  &           &  & & & &&(3)\\ 
V380 Ori        &$<$0.1  &           &  & & &&& (3)\\ 
HD 250550       &$<$0.12  &           &  & &&&& (3) \\ 
Lk H$\alpha$ 208&$<$0.03  &           &  & &&&& (3) \\ 
VV Ser          &$<$0.05  &           &  & &&&& (3) \\ 
WW Vul          &$<$0.09  &           &  & &&&& (3) \\ 
BD+46$^{\circ}$3471&$<$0.1  &           &  & &&&& (3) \\ 
Lk H$\alpha$ 233 &$<$0.04 &           &  & &&&& (3) \\ 
\hline
WL 16 & 0.17(0.03) &  & & & 5(0.4)&& & (4)\\
\hline
\hline
\end{tabular}

Integrated fluxes (after continuum substraction) and uncertainty (in between brackets) are in units of 10$^{-14}$ W/m$^2$.
For non-detections, we give for the 3.3, 6.2 and 11.3 $\mu$m feature 5$\sigma$ limits 
estimated from the spectrum of \cite{meeus2001} and assuming constant feature widths
taken from \cite{li2001a}. We have not included the 11.2(3) $\mu$m band
for the sources with strong silicate emission. \\
$^a$ From \cite{brooke93}. \\ 
$^{b}$ In the spectrum of HD 100546 a 11.2(3) $\mu$m PAH feature is present
but may be blended with the crystalline olivine band at 11.3 $\mu$m \cite[]{malfait98}.\\
$^c$ Measured in ground-based spectrum \cite[]{geballe97}. \\ 
$^{d}$ \cite{hanner94} report the presence of an 11.2 $\mu$m feature in their UKIRT spectrum.
They attribute this feature to the presence of crystalline silicate but it could also be a PAH band.\\
References: (1) \cite{vankerckhoven2002}; (2) \cite{meeus2001}; (3) \cite{brooke93};
(4) From the spectrum of \cite{tokunaga91} and \cite{hanner92} for the 3.3 and 11.3 $\mu$m feature, respectively.
\end{table*}

\newpage
\newpage

\begin{table*}[h!] \caption{Predicted strength of the PAH emission features integrated over the disk.}
\label{tab_mod_1}
\begin{tabular}{llllllll}
\hline
\hline
Model & 3.3 & 6.2 & 7.7 & 8.6 & 11.3 & 12.7 & 16.4\\
   & $[\mu m]$ & $[\mu m]$ & $[\mu m]$ & $[\mu m]$ & $[\mu m]$ & $[\mu m]$ & $[\mu m]$ \\
\hline
template    & 2.8 & 5.4 & 10.2 & 1.3 & 2.5 & 0.6 & 2.6  \\  
\hline
Ionized & 0.1 & 7.5 & 18.2 &2.25 & 0.9 & 0.13 & 0.8 \\
$f_H$=0.5 & 1.4 & 5.9 & 11 & 1 & 1.5 & 0.3  & 2.6\\
$N_C$=40  & 5.3 & 2.5 & 4.1 & 1.1 & 10.8 & 2.8 & 1 \\
$N_C$=40 - Ionized & 0.3 & 6.1 & 12 &3 & 3.9 &0.9 & 0.4 \\
\hline
\hline
\end{tabular}

Integrated fluxes (after continuum substraction) are in units of 10$^{-14}$ W/m$^2$.\\
\end{table*}

\end{document}